\documentclass[twocolumn,aps,prb,showpacs,preprintnumbers]{revtex4}%
\usepackage[latin9]{inputenc}
\usepackage{amsmath}
\usepackage{amssymb}
\usepackage{graphicx}
\usepackage{mathrsfs}
\usepackage{amsfonts}
\usepackage{amsthm}
\usepackage{color}
\usepackage{txfonts}%
\providecommand{\U}[1]{\protect\rule{.1in}{.1in}}
\makeatletter
\@ifundefined{textcolor}{}
{
\definecolor{BLACK}{gray}{0}
\definecolor{WHITE}{gray}{1}
\definecolor{RED}{rgb}{1,0,0}
\definecolor{GREEN}{rgb}{0,1,0}
\definecolor{BLUE}{rgb}{0,0,1}
\definecolor{CYAN}{cmyk}{1,0,0,0}
\definecolor{MAGENTA}{cmyk}{0,1,0,0}
\definecolor{YELLOW}{cmyk}{0,0,1,0}
}
\makeatother
\begin{document}
\title{Energy repartition in the nonequilibrium steady state}
\author{Peng Yan$^{1}$}
\author{Gerrit E.W. Bauer$^{2,3}$}
\author{Huaiwu Zhang$^{1}$}
\affiliation{$^{1}$School of Microelectronics and Solid-State Electronics and State Key
Laboratory of Electronic Thin Film and Integrated Devices, University of
Electronic Science and Technology of China, Chengdu 610054, China}
\affiliation{$^{2}$Institute for Materials Research and WPI-AIMR, Tohoku University, Sendai
980-8577, Japan}
\affiliation{$^{3}$Kavli Institute of NanoScience, Delft University of Technology,
Lorentzweg 1, 2628 CJ Delft, The Netherlands}

\begin{abstract}
The concept of temperature in non-equilibrium thermodynamcis is an outstanding
theoretical issue. We propose an energy repartition principle that leads to a
spectral (mode-dependent) temperature in steady state non-equilibrium systems.
The general concepts are illustrated by analytic solutions of the classical
Heisenberg spin chain connected to Langevin heat reservoirs with arbitrary
temperature profiles. Gradients of external magnetic fields are shown to
localize spin waves in a Wannier-Zeemann fashion, while magnon interactions
renormalize the spectral temperature. Our generic results are applicable to
other thermodynamic systems such as Newtonian liquids, elastic solids, and
Josephson junctions.

\end{abstract}

\maketitle

\section{Introduction}

Equilibrium thermodynamics provides a unified description of the macroscopic
properties of matter and its response to weak stimuli without referring to
microscopic mechanisms. Statistical mechanics, on the other hand, proceeds
from quantum mechanics and describes macroscopic observables in terms of
probabilities and averages. The combination of both fields leads to an
understanding of many physical and chemical phenomena from first principles.
Temperature is a principal quantity in the study of equilibrium physics.
Energy equipartition in classical equilibrium thermodynamics implies that
every quadratic normal mode \cite{Huang} carries on average an energy
$k_{B}T/2\ $(quantum statistics can be diregarded when mode energies are small
compared to $k_{B}T$) \cite{Tolman}. Here $k_{B}$ is the Boltzmann constant
and $T$ is the absolute temperature. The system temperature of a given system
can be obtained by, e.g., the kinetic approach \cite{Huang}, the entropy
method \cite{Callen}, and dynamical systems theory \cite{Rugh}.

In recent years the physics of nonequilibrium systems has attracted attention
from widely different disciplines, such as stochastic thermodynamics
\cite{Speck}, many-body localizations \cite{Huse}, and spin caloritronics
\cite{Gerrit}. One outstanding issue is the concept and proper definition of
the temperature of a nonequilibrium system. Most common is the local thermal
equilibrium approximation, assuming that spatially separated components of a
system thermalize with their immediate surroundings, while the global state of
the system is out of equilibrium. The spatially distributed local temperature
forms a spatial field that gives a good impression of the nonequilibrium
dynamics of the full system. This approach, however, often leads to
contradictions: the kinetic temperature has been found to differ from the
entropic temperature \cite{Narayanan}. This is no issue in equilibrium
systems, in which the temperature is constant and all modes in momentum space
share the same temperature.

Recently, the (equilibrium) thermodynamic entropy has been identified as a
Noether invariant associated with an infinitesimal nonuniform time translation
\cite{Sasa}. In nonequilibrium systems, however, the translational symmetry is
broken, so the entropy appears to be not well defined either.

In this work, we propose the principle of energy repartition in nonequilibrium
systems. It provides partial answers to these fundamental questions by
enabling us to define a spectral (mode-dependent) temperature \cite{Alicki}.
We illustrate the principle for magnons in a classical Heisenberg spin chain
connected to Langevin heat reservoirs with arbitrary temperature profiles. We
analytically solve the non-Markovian Landau-Lifshitz-Miyazaki-Seki (LLMS)
equation \cite{Miyazaki} {[}Eq. (\ref{LLMS}) below{]}, and find that the
steady-state non-equilibrium properties are governed by a set of normal-mode
temperatures that depend on the bath temperature profile, the boundary
conditions, and the ratio between the field gradient and the exchange coupling
between spins. We show that gradients of external magnetic fields localize
spin waves in the Wannier-Zeeman fashion, while weak many-body interactions
(nonlinearities) lead to a mode-temperature renormalization. The LLMS equation
encompasses \emph{all} standard equations for classical spin dynamics,
reducing to the (stochastic) Landau-Lifshitz-Gilbert (LLG) equation
\cite{Brown,Kubo,Palacios,Gilbert} and the Bloch equation \cite{Kawabata} in
respective limits. Our generic results should be widely applicable to describe
the semiclassical dynamics of other thermodynamic systems such as Newtonian
liquids, elastic solids, and Josephson junctions.

This paper is organized as follows: In Sec. II, the theoretical model is
presented. Section III gives the results and discussions: we derive the the
analytical solution of non-Markovian spin waves and propose the principle of
energy repartition in Sec. III A; temperature and chemical potential of
nonequilibrium magnons are calculated in Sec. III B; spin pumping and spin
Seebeck effects are analyzed in Sec. III C; Wannier-Zeeman localization due to
inhomogeneous magnetic fields and its effect on magnon transport are predicted
in Sec. III D; magnon-magnon interactions are perturbatively treated in Sec.
III E. Section IV is the summary.

\begin{figure}[ptbh]
\begin{centering}
\includegraphics[width=8cm]{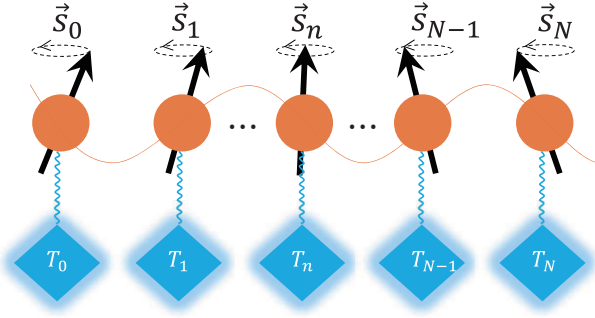}
\par\end{centering}
\caption{(Color online) Schematic of a monatomic spin chain consisting of
$N+1$ local magnetic moments $\vec{s}_{n}$ coupled with external Langevin bath
at temperature $T_{n}$, respectively, with $n=0,1,\cdots,N.$}%
\end{figure}

\section{Model}

We consider a classical monatomic spin chain along the $x$-direction,
consisting of $N+1$ local magnetic moments $\vec{S}_{n}=S\vec{s}_{n},$ where
the unit-vector $\vec{s}_{n}$ is the local spin direction, $S$ the total spin
per site, and $n=0,1,\cdots,N$. Each spin is in contact with a local Langevin
bath at temperature $T_{n}$, as shown in Fig. 1. Long wave-length excitations
of complex magnets such as yttrium iron garnet (YIG) can be treated by such a
model by coarse graining, i.e., letting each spin represents the magnetization
of a unit cell. Artificially fabricated exchange-coupled atomic spins on a
substrate \cite{Roland} is another physical realization of this model. The
magnetization dynamics can be described by the so-called
Landau-Lifshitz-Miyazaki-Seki (LLMS) equations \cite{Miyazaki}
\begin{equation}
\frac{d\vec{s}_{n}}{dt}=-\vec{s}_{n}\times\left(  \vec{H}_{\mathrm{eff}}%
+\vec{h}_{n}\right)  ,\frac{d\vec{h}_{n}}{dt}=-\frac{1}{\tau_{c}}\left(
\vec{h}_{n}-\chi\vec{s}_{n}\right)  +\vec{R}_{n}, \label{LLMS}%
\end{equation}
where $\vec{h}_{n}$ is the fluctuating magnetic field, $\vec{H}_{\mathrm{eff}%
}=\vec{H}_{n}+Ds_{n}^{z}\vec{z}+J\left(  \vec{s}_{n-1}+\vec{s}_{n+1}\right)
+\vec{H}_{d}$ is the effective field consisting of the external magnetic field
$\vec{H}_{n}$ and uniaxial anisotropy field with constant $D$ along the same
(here $z$-) direction, and the exchange constant $J$ initially taken to be
ferromagnetic, i.e., $J>0.$ $\vec{H}_{d}$ is caused by by long-range dipolar
fields, but is disregarded in the following. $\vec{R}_{n}$ is the random force
with zero average and a time-correlation function that satisfies the
fluctuation-dissipation theorem (FDT) \cite{Welton}:
\begin{equation}
\left\langle R_{n}^{i}\left(  t\right)  R_{n^{\prime}}^{j}\left(  t^{\prime
}\right)  \right\rangle =\left(  2\chi k_{B}T_{n}/\tau_{c}\right)
\delta_{nn^{\prime}}\delta_{ij}\delta\left(  t-t^{\prime}\right)  ,
\end{equation}
where $i,j=x,y,z$, the parameter $\chi$ describes the spin-bath coupling, and
$\tau_{c}$ is the relaxation time. In the following $H_{n}$, $D$, $J$, $H_{d}%
$, $h_{n}$, and $k_{B}T_{n}$ are all measured in Hz\textit{. }Equation
(\ref{LLMS}) has been very successful in atomistic simulations of ultrafast
spin dynamics for constant bath temperatures \cite{Atxitia} and can be derived
from microscopic spin-lattice or spin-electron couplings \cite{Atxitia,Peter}.
Here we introduce a spatially inhomogeneous thermal bath with arbitrary
temperature profiles. We assume statistical independence of neigboring baths,
i.e., a correlation length between reservoirs is shorter than the
(course-grained) lattice constant. By eliminating the fluctuating field
$\vec{h}_{n}$ in Eq. (\ref{LLMS}), we arrive at the following stochastic LLMS
with non-Markovian damping,
\begin{equation}
\frac{d\vec{s}_{n}}{dt}=-\vec{s}_{n}\times\left(  \vec{H}_{\mathrm{eff}}%
+\vec{\eta}_{n}\right)  +\chi\vec{s}_{n}\times\int_{-\infty}^{t}dt^{\prime
}\kappa\left(  t-t^{\prime}\right)  \frac{d\vec{s}_{n}\left(  t^{\prime
}\right)  }{dt^{\prime}}, \label{NMarkovian}%
\end{equation}
and a new stochastic field
\begin{equation}
\vec{\eta}_{n}=\int_{-\infty}^{t}dt^{\prime}\kappa\left(  t-t^{\prime}\right)
\vec{R}_{n}\left(  t^{\prime}\right)
\end{equation}
that is correlated as
\begin{equation}
\left\langle \eta_{n}^{i}\left(  t\right)  \eta_{n^{\prime}}^{j}\left(
t^{\prime}\right)  \right\rangle =\chi k_{B}T_{n}\delta_{nn^{\prime}}%
\delta_{ij}\kappa\left(  \left\vert t-t^{\prime}\right\vert \right)  ,
\end{equation}
with memory kernel $\kappa\left(  \tau\right)  =\exp\left(  -\tau/\tau
_{c}\right)  .$ Equation (\ref{NMarkovian}) is genuinely non-Markovian and has
been believed to be analytically intractable \cite{Miyazaki,Atxitia}.
Nevertheless, here we present an analytical solution for non-Markovian spin
waves, to the best of our knowledge for the first time.

\section{RESULTS AND DISCUSSIONS}

\subsection{Linear spin-wave theory}

For small-angle dynamics $\vec{s}_{n}\doteq\vec{z}+\left(  s_{n}^{x}\vec
{x}+s_{n}^{y}\vec{y}\right)  \ $with $\left\vert s_{n}^{x,y}\right\vert \ll1$
the stochastic LLMS equation reduces to
\begin{equation}
i\frac{d\psi_{n}}{dt}+\chi\int_{-\infty}^{t}dt^{\prime}\kappa\left(
t-t^{\prime}\right)  \frac{d\psi_{n}\left(  t^{\prime}\right)  }{dt^{\prime}%
}=-\sum_{m=0}^{N}\left(  JQ_{nm}+H_{n}\delta_{nm}\right)  \psi_{m}+\eta
_{n}\left(  t\right)  , \label{Schroedinger}%
\end{equation}
for the complex scalar-fields $\psi_{n}\left(  t\right)  =s_{n}^{x}+is_{n}%
^{y}$ and $\eta_{n}\left(  t\right)  =\eta_{n}^{x}+i\eta_{n}^{y},$ which are
correlated as
\begin{equation}
\left\langle \eta_{n}^{\ast}\left(  t\right)  \eta_{n^{\prime}}\left(
t^{\prime}\right)  \right\rangle =2\chi k_{B}T_{n}\delta_{nn^{\prime}}%
\delta_{ij}\kappa\left(  \left\vert t-t^{\prime}\right\vert \right)  ,
\end{equation}
where $^{\ast}$ is the complex conjugate. The extra factor 2 reflects energy
equipartition since $\eta_{n}$ incorporates two degrees of freedom. $Q$ is a
$\left(  N+1\right)  \times\left(  N+1\right)  $ symmetric quasi-uniform
tridiagonal canonical matrix that does not depend on material parameters (see
Appendix A). In $H_{n}=H+\varepsilon n,\ \varepsilon$ models external or
anisotropy field gradients \cite{Antropov,Sukhov}. Since in general, matrices
$Q$ and diag$\left\{  H_{n}\right\}  $ cannot be diagonalized simultaneously,
we introduce a new matrix $\check{Q}=Q+\left(  \varepsilon/J\right)
\mathrm{diag}\left\{  n\right\}  $ that satisfies $JQ_{nm}+H_{n}\delta
_{nm}=J\check{Q}_{nm}+H\delta_{nm}.$ We remove the integral in Eq.
(\ref{Schroedinger}) by taking the time-derivative%
\begin{align}
&  i\frac{d^{2}\psi_{n}}{dt^{2}}+\sum_{m=0}^{N}\left[  J\check{Q}_{nm}+\left(
H+\chi+i\tau_{c}^{-1}\right)  \delta_{nm}\right]  \frac{d\psi_{n}}%
{dt}\nonumber\\
&  =-\tau_{c}^{-1}\sum_{m=0}^{N}\left(  J\check{Q}_{nm}+H\delta_{nm}\right)
\psi_{m}+R_{n}\left(  t\right)  ,
\end{align}
where $R_{n}\left(  t\right)  =R_{n}^{x}+iR_{n}^{y}$ is correlated as
\begin{equation}
\left\langle R_{n}^{\ast}\left(  t\right)  R_{n^{\prime}}\left(  t^{\prime
}\right)  \right\rangle =\left(  4\chi k_{B}T_{n}/\tau_{c}\right)
\delta_{nn^{\prime}}\delta\left(  t-t^{\prime}\right)  .
\end{equation}
In the limit of $\tau_{c}\rightarrow0,$ the above equation reduces to the
Markovian LLG:
\begin{equation}
\left(  i+\alpha\right)  \frac{d\psi_{n}}{dt}=-\sum_{m=0}^{N}\left(
J\breve{Q}_{nm}+H\delta_{nm}\right)  \psi_{m}+\xi_{n}\left(  t\right)  ,
\end{equation}
with correlator
\begin{equation}
\left\langle \xi_{n}^{\ast}\left(  t\right)  \xi_{n^{\prime}}\left(
t^{\prime}\right)  \right\rangle =4\alpha k_{B}T_{n}\delta_{nn^{\prime}}%
\delta\left(  t-t^{\prime}\right)
\end{equation}
expressed in terms of the Gilbert damping constat $\alpha=\chi\tau_{c}.$ The
mathematical structure is identical to that of fluctuating heat \cite{Zarate}
and/or mass \cite{Tremblay} transport and the widely studied macroscopic
fluctuation theory of fluids \cite{Lorenzo}, where the scalar field $\psi$
represents temperature \cite{Zarate} or number density fluctuations
\cite{Tremblay}, while $\xi\left(  t\right)  $ is the divergence of a heat or
particle current.

The symmetric tridiagonal matrix $\check{Q}$ can be diagonalized by a linear
transformation $P^{-1}\check{Q}P$ with an orthogonal matrix $P$ which solely
depends on the ratio $\varepsilon/J$. This is equivalent to an expansion of
the field into normal magnon modes $\phi_{k}=\sum_{n=0}^{N}P_{kn}^{-1}\psi
_{n}$ that obey
\begin{equation}
\frac{d^{2}\phi_{k}}{dt^{2}}+\nu_{k}\frac{d\phi_{k}}{dt}-\frac{i\omega_{k}%
}{\tau_{c}}\phi_{k}=f_{k}\left(  t\right)  , \label{NewStochastic}%
\end{equation}
where $\omega_{k}=H+J\lambda_{k}$ is the eigenfrequency of the $k$-th mode,
$\lambda_{k}$ is the $k$-th eigenvalue of $\check{Q},$ and $\nu_{k}=\tau
_{c}^{-1}-i\left(  \chi+\omega_{k}\right)  $. The structure of Eq.
(\ref{NewStochastic}) is reminiscent of the thermal acoustic wave equations
\cite{Huang} and the dynamic equations of fluctuating superconducting
Josephson junctions \cite{Raghavan}. The boundary conditions affect the
dispersion relation $\omega_{k}$. The modes interact via the transformed
stochastic variable $f_{k}=-i\sum_{n=0}^{N}P_{kn}^{-1}R_{n}\ $with non-local
correlator
\begin{equation}
\left\langle f_{k}^{\ast}\left(  t\right)  f_{k^{\prime}}\left(  t^{\prime
}\right)  \right\rangle =\left(  4\chi k_{B}\mathcal{T}_{kk^{\prime}}/\tau
_{c}\right)  \delta\left(  t-t^{\prime}\right)  , \label{NewCorrelation}%
\end{equation}
introducing the temperature matrix
\begin{equation}
\mathcal{T}_{kk^{\prime}}=\sum_{n=0}^{N}P_{nk}P_{nk^{\prime}}T_{n}.
\label{TemperatureMatrix}%
\end{equation}
$\mathcal{T}$ is diagonal in the absence of temperature gradients, i.e., when
$T_{n}=T$ $\forall n.$

We now show that the diagonal terms $\mathcal{T}_{kk}$ encode the energy
distribution over the different magnon modes in the nonequilibrium steady
state. The average energy of the $k$-th magnon mode is $E_{k}=\omega
_{k}\left\langle \phi_{k}^{\ast}\phi_{k}\right\rangle /2,$ where the
expectation value $\left\langle \cdots\right\rangle $ is taken over different
realizations of the thermal noise $R_{n}\left(  t\right)  $ and $\left\langle
\phi_{k}^{\ast}\phi_{k}\right\rangle /2$ is the magnon number. Equation
\eqref{NewStochastic} can be solved exactly by introducing the Green function
corresponding to the left-hand side and integrating over the noise source
term:
\begin{equation}
\phi_{k}\left(  t\right)  =\int_{-\infty}^{t}dt^{\prime}\frac{1}{c_{1}-c_{2}%
}\left[  e^{-c_{2}\left(  t-t^{\prime}\right)  }-e^{-c_{1}\left(  t-t^{\prime
}\right)  }\right]  f_{k}\left(  t^{\prime}\right)  , \label{Solution}%
\end{equation}
with two complex numbers
\begin{equation}
c_{1,2}=\left(  \nu_{k}\pm\sqrt{\nu_{k}^{2}+4i\tau_{c}^{-1}\omega_{k}}\right)
/2.
\end{equation}
We thus arrive at the central result of this work that the energy stored in
mode $k$ is nothing but the thermal energy as defined by the diagonal elements
of $\mathcal{T}$:
\begin{equation}
E_{k}=k_{B}\mathcal{T}_{kk}. \label{Repartition}%
\end{equation}
The entropy of the nonequilibrium steady system then can be expressed as
$\mathcal{S}=-k_{B}\sum_{k}p_{k}\ln p_{k},$with the probability distribution
$p_{k}=\left\langle \left\vert \phi_{k}\right\vert ^{2}\right\rangle
/\sum_{k^{\prime}}\left\langle \left\vert \phi_{k^{\prime}}\right\vert
^{2}\right\rangle .$ Interestingly, for homogeneous external magnetic fields
$\mathcal{T}_{kk}$ is parameter-free, depending only on the bath temperature
profile $T_{n}$ and the boundary conditions. A magnetic field gradient
modifies the mode temperature only via the ratio $\varepsilon/J.$ The memory
kernel with relaxation time $\tau_{c}$ does not affect the repartition.
Although we consider an exponential memory kernel here, we envision that the
obtained energy repartition principle \eqref{Repartition} should be robust to
the specific form of the kernels. The generalization to two spins in the unit
cell leads to acoustic and optical magnon branches and can be used to study
ferrimagnets and antiferromagnets \cite{Joe}. In the following, we limit
ourselves to the temperature distribution of non-equilibrium ferromagnetic
magnons. Off-diagonal terms $\mathcal{T}_{kk^{\prime}}$ $\left(  k\neq
k^{\prime}\right)  $ encode the magnonic spin current which can be obtained
from the spin continuity equation \cite{Zotos,Jencic}
\begin{equation}
\vec{j}_{M,n}=J\vec{s}_{n-1}\times\vec{s}_{n},\left(  0<n\leqslant N\right)  .
\end{equation}
Its DC component can be expanded into normal modes as
\begin{equation}
j_{M,n}^{z}=J\sum_{kk^{\prime}}P_{\left(  n-1\right)  k}P_{nk^{\prime}%
}\operatorname{Im}\left\langle \phi_{k}^{\ast}\phi_{k^{\prime}}\right\rangle ,
\end{equation}
where $\operatorname{Im}\cdots$ denotes the imaginary part. The associated
real space \textit{magnon density }distribution \cite{ShufengZhangprl}
$\rho_{M,n}=\left\langle \psi_{n}^{\ast}\psi_{n}\right\rangle /2$ is conjugate
to the magnon number in reciprocal space\textit{ }$\left\langle \phi_{k}%
^{\ast}\phi_{k}\right\rangle /2.$ These quantities are expressed in terms of
spectral temperatures in Appendix A.

\subsection{Temperature and chemical potential of non-equilibrium magnons
under uniform magnetic field}

We first consider a simple case with a vanishing field gradient $\left(
\varepsilon=0\right)  .$ Under free boundaries (no pinning), we derive
(Appendix A)
\begin{equation}
\mathcal{T}_{kk}=\left\{
\begin{array}
[c]{c}%
\bar{T},\text{ \ \ \ \ \ \ \ \ \ \ \ \ \ \ \ \ \ \ \ \ \ \ \ \ \ }k=0,\\
\bar{T}+\sum_{n=0}^{N}\frac{T_{n}}{N+1}\cos\frac{\left(  2n+1\right)  k\pi
}{N+1},\text{ \ }k\neq0,
\end{array}
\right.  \label{Free}%
\end{equation}
where $\bar{T}=\sum_{n=0}^{N}T_{n}/\left(  N+1\right)  $ is the average bath
temperature. The energy stored in mode $k$ emerges as a correction to the
average temperature $\bar{T},$ but never exceeds $\pm\bar{T}.$ $\mathcal{T}%
_{kk}-\bar{T}$ is an average over the bath temperature profile weighted by a
cosine function. We study the spectrally resolved temperature $\mathcal{T}%
_{kk}$ for five different model baths, all with $T_{0}=300,T_{N}=350,$ and
$N=99$ [see Fig. 2(a)] (in arbitrary temperature units): (i) a linear
temperature profile, i.e., $T_{n}=T_{0}+\left(  T_{N}-T_{0}\right)  n/N$, (ii)
a quadratic profile, i.e., $T_{n}=T_{0}+\left(  T_{N}-T_{0}\right)  \left(
n/N\right)  ^{2}$, (iii) a \textquotedblleft subduplicate\textquotedblright%
\ profile, i.e., $T_{n}=T_{0}+\left(  T_{N}-T_{0}\right)  \sqrt{n/N}$, (iv) a
Sanders-Walton profile, i.e.,
\begin{equation}
T_{n}=T_{0}+\frac{T_{N}-T_{0}}{N+2\mu\sinh\left(  \frac{N}{\nu}\right)
}\left[  n+\mu\left(  \sinh\frac{2n-N}{\nu}+\sinh\frac{N}{\nu}\right)
\right]
\end{equation}
with adjustable parameters $\mu$ and $\nu$ \cite{Sanders,Xiao,Konstantin}
chosen to be $\mu=1$ and $\nu=16$, and (v) an asymmetric Heaviside step
function \cite{Ritzmann} at $10+\left(  N+1\right)  /2$. While a linear and
sinh profiles can make physical sense being solutions of a simple heat
diffusion equation, arbitrary temperature profiles can be engineered in terms
of a string of heat sources such as Peltier cells placed along the spin chain.

\begin{figure}[ptbh]
\begin{centering}
\includegraphics[width=8cm]{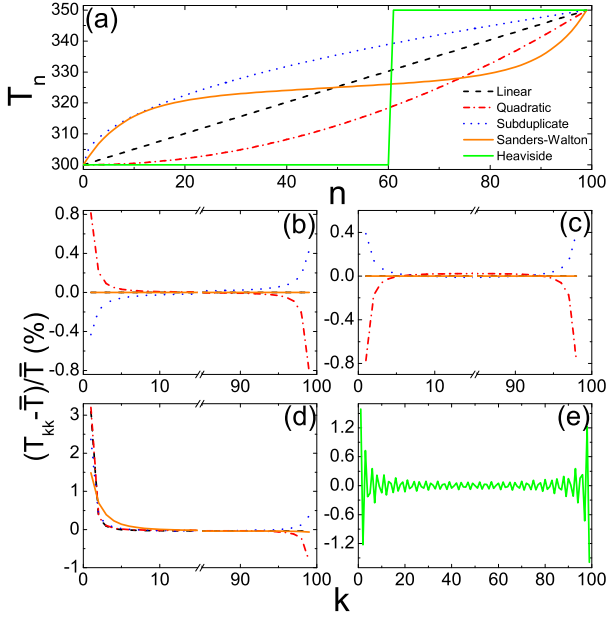}
\par\end{centering}
\caption{(Color online) (a) Thermal bath temperature profiles chosen to study
the mode-resolved temperature of nonequilibrium magnons. (b)-(d) Dependence of
the temperatures of normal magnon modes $\phi_{k}$ on boundary conditions: (b)
Both ends are free, (c) Both ends are pinned, and (d) The left end is pinned
while the right one is free. (e) Temperature of $k$-magnons under a asymmetric
Heaviside temperature distribution with free boundary conditions. The applied
magnetic field is uniform.}%
\end{figure}

Figure 2(b)\textit{ }shows the resulting $\mathcal{T}_{kk}$ for free boundary
conditions. The magnon temperature does not deviate from the average
temperature $\bar{T}$ for both the linear and the Sanders-Walton profile. The
correction terms in Eq. \eqref{TemperatureMatrix} vanish for all temperature
profiles that are odd around $\left(  N/2,\bar{T}\right)  .$ For free boundary
conditions the equipartition at equilibrium persists for temperature profiles
with odd symmetry. For quadratic (subduplicate) profiles, on the other hand,
low- (high-) $k$ magnons are heated and high- (low-) $k$ magnons cooled. In
general, pinning can reduce the magnon amplitude at the sample boundaries,
which obviously affects transport. However, boundary conditions also modify
the energy repartition of non-equilibrium magnons, as demonstrated in Fig.
2(c) for fixed (pinned) boundary conditions (the analytical expression of
$\mathcal{T}_{kk}$ are given in Appendix A). Notably, long-wavelength magnons
are strongly affected by the boundary conditions, which leads to the inverted
temperature profile when magnons are pinned and thereby do not sense the
temperature at the edges. Figure 2(d) shows $\mathcal{T}_{kk}$ as a function
of $k$ under boundary conditions with a pinned left and a free right terminal.
Since the boundaries now break symmetry, even for the antisymmetric profiles
the magnon temperature become distributed; the low-$k$ magnons are getting
hotter. We find that a higher asymmetry of either the bath temperature profile
or the boundary condition leads to a smaller decay length in the reciprocal
space ($k$ space). Figure 2(e) shows oscillations of the mode-dependent
temperatures for a non-symmetric and non-adiabatic thermal bath profile, i.e.,
with a Heaviside step function displaced from the midpoint. Though calculated
for free boundary conditions this feature is robust with respect to other choices.

\begin{figure}[ptbh]
\begin{center}
\includegraphics[width=8cm]{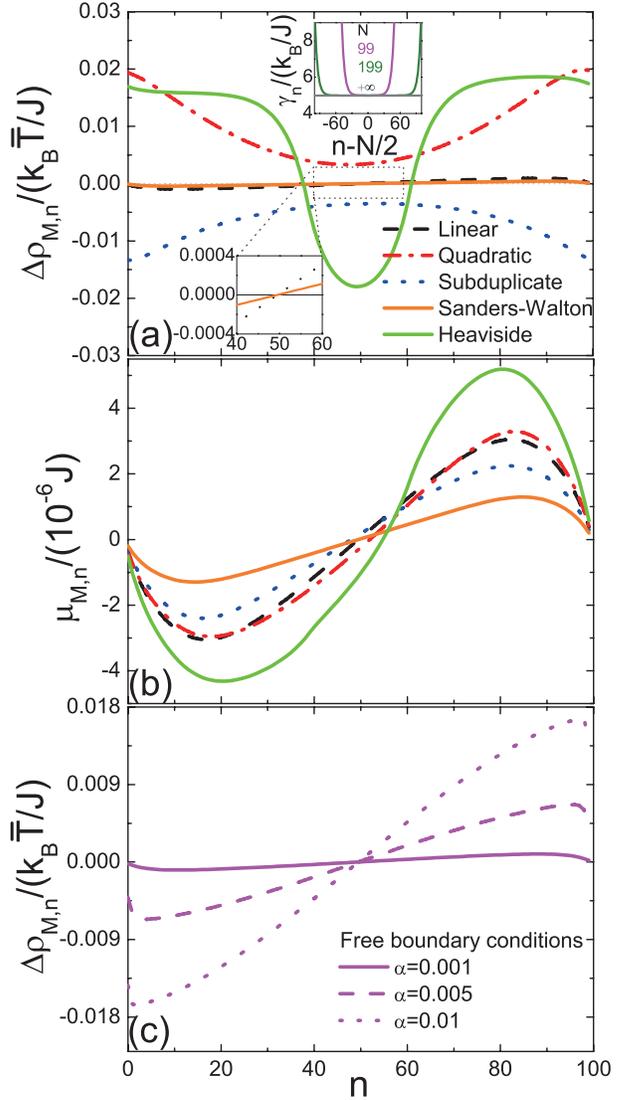}
\end{center}
\caption{(Color online) (a) Spatial distribution of thermally induced magnon
accumulations for different heat-bath profiles. Inset (upper-middle):
$\gamma_{n}$ as function of system size $N$. Inset (lower-left corner): Zoom
of the accumulation for linear and Sanders-Walton bath profiles at the sample
center. (b) Magnon chemical potential distribution for different heat baths.
In (a) and (b), we set damping parameter $\alpha=0.001.$ (c) Magnon
accumulation as a function of the damping parameter for a linear heat-bath. In
calculations, we consider free boundary conditions at the edges and set
$H/J=0.01$.}%
\end{figure}

For free boundary conditions and bath temperature profiles with odd symmetry
with respect to $\left(  N/2,\bar{T}\right)  \ $all magnons share the same
temperature $\bar{T},$ cf. Eq. \eqref{Free}. One might therefore naively
conclude that the magnon distribution is then not modified by the temperature
gradient. However, the local temperature differences between bath and magnon
would make the steady state unsustainable since we find a heat current-induced
\textit{magnon accumulation} $\Delta\rho_{M,n}=\rho_{M,n}-\gamma_{n}\bar{T}$
with $\gamma_{n}=\sum_{k}\left(  P_{nk}\right)  ^{2}k_{B}/\omega_{k}$
(Appendix A). Figure 3(a) shows the calculated spatial distribution
$\Delta\rho_{M,n}$ for different heat baths and free boundary conditions. For
lattice temperatures with odd-symmetry, the magnon accumulation around the
center $N/2$ increases linearly with site $n$ {[}the lower-left-corner inset
of Fig. 3(a) zooms in on the details{]} with a slope that depends on the shape
of the temperature profile. The magnon accumulation is distributed in space,
in spite of the uniform magnon temperature $\mathcal{T}_{kk}=\bar{T}$ $\forall
k$ at all sites $n.$ Therefore, the magnon distribution cannot be
parameterized by temperature alone. The solution is provided by introducing a
distributed magnon chemical potential. A finite magnon chemical potential is
the precursor of the magnon Bose-Einstein (or Rayleigh-Jeans) condensation
that has been observed in magnetic insulators parametrically pumped by
microwaves \cite{Peter}.

The semiclassical nonequilibrium distribution function of magnons can be
described by Bose-Einstein statistics
\begin{equation}
f_{\mathrm{BE}}\left(  k,n\right)  =\frac{1}{\exp\left(  \frac{\omega_{k}%
-\mu_{M,n}}{k_{B}\mathcal{T}_{kk}}\right)  -1}%
\end{equation}
in \emph{phase space} spanned by coordinate and momentum, which in the
high-temperature limit approaches the Rayleigh-Jeans distribution
$f_{\mathrm{BE}}\left(  k,n\right)  \rightarrow k_{B}\mathcal{T}_{kk}/\left(
\omega_{k}-\mu_{M,n}\right)  .$\textit{ }The magnon chemical potential profile
$\mu_{M,n}$ can therefore be determined by equating
\begin{equation}
\rho_{M,n}=\sum_{k}\left(  P_{nk}\right)  ^{2}\frac{k_{B}\mathcal{T}_{kk}%
}{\omega_{k}-\mu_{M,n}},
\end{equation}
with $\left\langle \psi_{n}^{\ast}\psi_{n}\right\rangle /2$.

The calculated $\mu_{M,n}$ for different heat baths under free boundary
conditions are shown in Fig. 3(b). At equilibrium $\mu_{M,n}$ vanishes and the
local magnon density is governed by the magnon temperature only. For
quadratic, subduplicate, and Heaviside profiles, the magnon accumulation is
non-monotonic. In a subduplicate bath, it first increases and then decreases
with $n$, opposite to the cases of quadratic and Heaviside profiles. We
therefore conclude that heat-bath temperature profiles can strongly affect the
magnon accumulation. In Fig. 3(c), by tuning the damping parameter $\alpha,$
we find that a larger dissipation causes a spatially steeper magnon
accumulation (a smaller diffusion length) under free boundary conditions.
Using other boundary conditions does not change the results qualitatively.

\begin{figure}[ptbh]
\begin{centering}
\includegraphics[width=8cm]{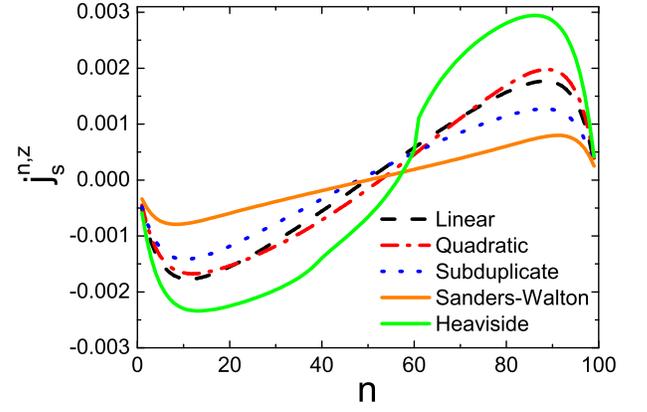}
\par\end{centering}
\caption{(Color online) Spin Seebeck spin current (in units of $\hbar
g_{\mathrm{eff}}^{\uparrow\downarrow}k_{B}\bar{T}/\pi\ $in a metal contact
attached to site $n$ for different heat-bath profiles and mixed boundary
conditions. Parameters used in the calculations are $\alpha=0.001$ and
$H/J=0.01.$ }%
\end{figure}

\subsection{Spin pumping and spin Seebeck effects}

Thermal spin currents can be detected by heavy normal metal contacts that
convert them into a transverse voltage by the inverse spin Hall effect
\cite{Uchida1}. We can model this situation by contacting the spin chain
either at the two ends or at some intermediate site. The former configuration
corresponds to the \textquotedblleft longitudinal\textquotedblright\ spin
Seebeck effect \cite{Uchida2,SHuang,Qu,Weiler,Ken,Andreas}, while the latter
one is referred to as \textquotedblleft transverse\textquotedblright%
\ \cite{Uchida1,Jaworski1,Jaworski2,Zink,Schmid,Meier} or \textquotedblleft
non-local\textquotedblright\ \cite{Shan}. The spin dynamics at the interface
pumps a spin current into the contact at site $n$ given by
\begin{equation}
\vec{j}_{s,n}=g_{\mathrm{eff}}^{\uparrow\downarrow}\frac{\hbar}{4\pi}\vec
{s}_{n}\times\frac{d\vec{s}_{n}}{dt},
\end{equation}
where $g_{\mathrm{eff}}^{\uparrow\downarrow}$ is the effective spin-mixing
conductance including a back-flow correction \cite{Hujun} and/or spin-orbit
coupling at the interface \cite{Kai}. Its averaged DC component reads
\begin{equation}
j_{s,n}^{z}=-g_{\mathrm{eff}}^{\uparrow\downarrow}\frac{\hbar}{4\pi}%
\sum_{k,k^{\prime}}P_{n,k}P_{n,k^{\prime}}\operatorname{Im}\left\langle
\dot{\phi}_{k}^{\ast}\phi_{k^{\prime}}\right\rangle .
\end{equation}
In the small dissipation/Markovian limit, the pumped DC spin current can be
expressed as
\begin{equation}
j_{s,n}^{z}=\frac{2\hbar g_{\mathrm{eff}}^{\uparrow\downarrow}}{\pi\left(
1+\alpha^{2}\right)  }\sum_{kk^{\prime}}P_{nk}P_{nk^{\prime}}k_{B}\left(
\mathcal{T}_{kk^{\prime}}-T_{e}\delta_{kk^{\prime}}\right)  \mathcal{G}\left(
\alpha,\omega_{k},\omega_{k^{\prime}}\right)  ,\label{TrueSC}%
\end{equation}
where
\begin{equation}
\mathcal{G}=\frac{\alpha^{2}\omega_{k}\omega_{k^{\prime}}}{\alpha^{2}\left(
\omega_{k}+\omega_{k^{\prime}}\right)  ^{2}+\left(  \omega_{k}-\omega
_{k^{\prime}}\right)  ^{2}}.
\end{equation}
Experimentally, this spin current can be detected by the inverse spin Hall
voltage in attached heavy metal contacts. Here we include the Johnson-Nyquist
noise generated in the metal that is proportional to the electron temperature
$T_{e}$, usually assumed to be in equilibrium with its phonon temperature.
Disregarding the Kapitza interface heat resistance, the phonon temperature is
continuous over the interface and $T_{e}=T_{n}$. For small damping,
$\alpha\simeq10^{-5}$ in YIG, the cross correlations between modes become
unimportant and
\begin{equation}
j_{s,n}^{z}\simeq g_{\mathrm{eff}}^{\uparrow\downarrow}\frac{\hbar}{2\pi}%
\sum_{k}\left(  P_{nk}\right)  ^{2}k_{B}\left(  \mathcal{T}_{kk}-T_{e}\right)
,
\end{equation}
as found in conventional spin Seebeck theory \cite{Xiao} for uniform magnon
temperature $\mathcal{T}_{kk}=T_{m}$ $\forall k$. According to this theory,
the spin Seebeck effect vanishes when magnon and electron temperatures are the
same. However, the full Eq. \eqref{TrueSC} reveals the limitations of this
approximation: the off-diagonal terms generate an SSE even in the absence of a
temperature difference between magnons and electrons. Figure 4 shows the
spatial distribution of the pumped spin current \eqref{TrueSC} for $T_{e}%
=\bar{T},$ i.e., the contribution to the SSE driven by the chemical potential
alone, for different bath temperature profiles and mixed boundary conditions.
The details of the bath profile strongly affect the distribution and magnitude
of the spin current and spin Seebeck effect.

\subsection{Wannier-Zeeman localization}

It follows from Eq. (\ref{Schroedinger}) that magnetic field gradients act on
magnons like electric fields act on electrons. Sufficiently strong electric
potential gradients in crystals can cause Wannier-Stark electron localization
\cite{Emin}. We may therefore expect an analogous Wannier-Zeeman magnon
localization in strongly inhomogeneous magnetic fields, which may modify the
mode temperature of magnons. The matrix $\check{Q}$ generally can in that
limit not be diagonalized analytically anymore, but small or a large
magnetic-field gradient can be treated perturbatively. In the limit of
\emph{large} magnetic field gradients $\left\vert \varepsilon/J\right\vert
\gg1$ and free boundary conditions: $\omega_{0}=H+J,$ $\omega_{N}%
=H+J+\varepsilon N,$ $\omega_{k}=H+2J+\varepsilon k$ for $0<k<N,$ and
$P_{nk}=\delta_{nk}.$ The spectrum then becomes a Wannier-Zeeman ladder. The
temperature matrix $\mathcal{T}_{kk^{\prime}}=\delta_{kk^{\prime}}T_{k}$ is
then diagonal even at nonequilibrium, i.e., the localization length is of the
order of the lattice constant. The magnon density becomes $\rho_{M,n}%
=k_{B}T_{n}/\omega_{n},$ thereby recovering the classical Rayleigh-Jeans
distribution with zero chemical potential, i.e. local thermal equilibrium.
Strong magnon localizations renders the spin chain insulating since
$j_{M,n}^{z}=0$. In the limit of small damping, the pumped spin current
becomes $j_{s,n}^{z}=g_{\mathrm{eff}}^{\uparrow\downarrow}\left(  \hbar
/2\pi\right)  k_{B}\left(  T_{n}-T_{e}\right)  ;$ the spin Seebeck effect
becomes local and vanishes when electrons on the metal side of the contact are
at the same temperature as the thermal bath (phonons) on the magnetic side.

Numerical calculations describe the transition from extended Bloch states for
small field-gradients to localized Wannier-Zeeman ladder states under large
magnetic field gradients (referring to Appendix A for details and figures).
The localization length $L=1/\sum_{n=0}^{N}\left(  P_{nk}\right)  ^{4}$ (in
units of the lattice constant) shrinks with increasing gradient, down to unity
in the limit of high field-gradients. The localized magnon states shift from
the low- to the high-field region with increasing energy. For a long chain
$\left(  N\rightarrow\infty\right)  ,$ we find an asymptotic $L\sim-1/\left[
\left(  \varepsilon/J\right)  \ln\left(  \varepsilon/J\right)  \right]  $ for
$\varepsilon/J\rightarrow0.$ Magnon localization suppresses the transverse or
non-local spin Seebeck effect. However, most experiments are carried out on
YIG films with very small anisotropy, which makes observation difficult. On
the other hand, strong perpendicular anisotropies can be induced by alloying
and doping (but preserving high magnetic quality) \cite{Deleeuw,Tsang,Coey}.
In (YBi)$_{3}$(FeGa)$_{5}$O$_{12}$ this is reflected by domain wall widths of
$8-11$ lattice constants \cite{Novoselov}. The material parameters at low
temperatures are \cite{Cherepanov,Kreisel,Novoselov} an exchange coupling
$J=1.29$ K and crystalline magnetic anisotropy $D=0.3$ K, and lattice constant
$a=1.24$ nm.\textit{ }An upper bound for the field gradient generated by a
position dependent magnetic anisotropy in a temperature gradient can be
obtained assuming its low temperature value on the cold side and a vanishing
one at the hot side, or $\varepsilon=\left(  D/l\right)  a=4\times10^{-7}$ K
and $\varepsilon/J=3\times10^{-7}.$ This leads to a magnon localization length
$L=-1/\left[  \left(  \varepsilon/J\right)  \ln\left(  \varepsilon/J\right)
\right]  \times a=0.3$ mm. When the magnons are localized on the scale of the
metal contact widths (typically 0.1 mm, see e.g. Ref. \onlinecite{Meier}, and
references therein) we predict a suppressed spin Seebeck signal. Magnon
localization can also be induced by applying magnetic field gradients, for
example by the stray fields of proximity ferromagnets or by the Oersted fields
due to current-carrying wires close to the magnon conduits. Magnetic write
heads generate local field gradients of up to 20 MT/m. Analogous to electronic
Wannier-Stark localizations in semiconductor superlattices \cite{Rossi},
magnonic crystals with tunable lattice periods can display magnon localization
at possibly much weaker inhomogeneous magnetic fields.

\subsection{Magnon-magnon interactions}

Results above assume the presence of magnon-phonon thermalization, but absence
of magnon-magnon interactions that modify the equations of motion for higher
magnon densities. Anisotropy-mediated magnon interactions dominate in the
long-wave lengths regime considered here \cite{Kosevich,Slavin,Simon}.
Adopting the Markov approximation and to leading order in the magnon density,
we arrive at a dissipative discrete nonlinear Schrödinger (DNLS) equation with
stochastic sources and a local interaction
\begin{equation}
\left(  i+\alpha\right)  \frac{d\psi_{n}}{dt}=-\sum_{m=0}^{N}\left[
J\breve{Q}_{nm}+\left(  H-\nu\left\vert \psi_{m}\right\vert ^{2}\right)
\delta_{nm}\right]  \psi_{m}+\xi_{n}\left(  t\right)  , \label{MB}%
\end{equation}
where $\nu$ is the interaction strength governed by the anisotropy constant
$D$ but treated here as a free parameter.\textit{ }For $\nu=0,$ eigenstates
are affected by magnetic field gradients $\varepsilon,$ as discussed above.
The mode frequency splitting $\Delta\omega\sim\min\left[  J\left(
\lambda_{k+1}-\lambda_{k}\right)  \right]  ,$ while for large $\varepsilon$,
$\Delta\omega\sim$ $\varepsilon.$ The nonlinearity in Eq. (\ref{MB}) for the
uniaxial anisotropy considered ($D,\nu>0$) corresponds to an attractive
interaction and a frequency red shift $\delta\omega_{n}\sim\nu\left\vert
\psi_{n}\right\vert ^{2}.$ The interaction is assumed short range, which is
allowed when dipolar coupling between spins is small in our coarse grained
model. We may then expect three qualitatively different regimes: (i)
$\left\vert \nu\right\vert <\Delta\omega;$ (ii) $\Delta\omega<\left\vert
\nu\right\vert <\Delta;$ (iii) $\Delta<\left\vert \nu\right\vert ,$ where the
band width $\Delta=\omega_{N}-\omega_{0}.$ In case (i), the local frequency
shift is smaller than the spacing $\Delta\omega$. Therefore, the long-time
dynamics is not modified from the limit $\nu=0$. For (ii) non-linearities
become important since the mode frequencies overlap. In the limit (iii) the
interaction is stronger than the non-interacting band width, drastically
transforming the spectrum. Discrete bound states may develop at the band
edges, leading to interaction induced self-trapping \cite{Raghavan}.

We may expand (\ref{MB}) into normal modes as before to obtain
\begin{equation}
\left(  i+\alpha\right)  \frac{d\phi_{k}}{dt}=-\omega_{k}\phi_{k}+\nu
\sum_{k_{1},k_{2},k_{3}}I_{k,k_{1},k_{2},k_{3}}\phi_{k_{1}}^{\ast}\phi_{k_{2}%
}\phi_{k_{3}}+\zeta_{k}\left(  t\right)  ,
\end{equation}
where the matrix elements
\begin{equation}
I_{k,k_{1},k_{2},k_{3}}=\sum_{n}P_{nk}P_{nk_{1}}P_{nk_{2}}P_{nk_{3}}%
\end{equation}
describe four-magnon scattering events and the stochastic variables are
correlated as
\begin{equation}
\left\langle \zeta_{k}^{\ast}\left(  t\right)  \zeta_{k^{\prime}}\left(
t^{\prime}\right)  \right\rangle =4\alpha k_{B}\mathcal{T}_{kk^{\prime}}%
\delta\left(  t-t^{\prime}\right)  .
\end{equation}
For arbitrary field gradients, we obtain the analytical formula of the
nonlinearity correction to the energy repartition up to the first-order of
$\nu$ as follows (Appendix B) \begin{widetext}
\begin{align}
k_{B}\mathcal{T}_{kk}^{\prime}=k_{B}\mathcal{T}_{kk}+16\nu\sum_{k_{1},k_{2},k_{3}}I_{k,k_{1},k_{2},k_{3}}\frac{\alpha^{2}\left(k_{B}\mathcal{T}_{kk_{3}}\right)\left(k_{B}\mathcal{T}_{k_{1}k_{2}}\right)\left[\left(-3+\alpha^{2}\right)\omega_{k}\omega_{k_{1}}+\left(1+\alpha^{2}\right)\left(\omega_{k}\omega_{k_{2}}+\omega_{k_{1}}\omega_{k_{3}}+\omega_{k_{2}}\omega_{k_{3}}\right)\right]}{\left[\left(\omega_{k_{1}}-\omega_{k_{2}}\right)^{2}+\alpha^{2}\left(\omega_{k_{1}}+\omega_{k_{2}}\right)^{2}\right]\left[\left(\omega_{k}-\omega_{k_{3}}\right)^{2}+\alpha^{2}\left(\omega_{k}+\omega_{k_{3}}\right)^{2}\right]},\label{Firstorder1}
\end{align}
\end{widetext}where we introduce the renormalized thermal energy
$k_{B}\mathcal{T}_{kk}^{\prime}=\omega_{k}\left\langle \phi_{k}^{\ast}\phi
_{k}\right\rangle /2.$ It reduces to $\mathcal{T}_{kk}^{\prime}=\left(
1+\Lambda\right)  \mathcal{T}_{kk}$ in the strongly localized limit in leading
order of the small parameter $\Lambda=4\nu k_{B}\mathcal{T}_{kk}/\omega
_{k}^{2}.$\textit{ }The interaction generates a red-shift of the spectrum and
corresponding higher thermal occupation, as confirmed by numerical simulations
for few-spin systems (Aappendixes C, D, and E) for both strong and relatively
weak localizations. The nonlinearity is therefore acting like an additional
heat source leading to mode-dependent corrections to the temperature that are
observable in the spin Seebeck effect, e.g. by tuning the anisotropy while
keeping other material parameters approximately constant.\emph{ }

\section{Summary}

To conclude, we report here a principle of energy repartition for
nonequilibrium system. We illustrate the general principle at the hand of
analytical solutions of the non-Markovian Landau-Lifshitz-Miyazaki-Seki
equations. We find that fluctuations are governed by a set of normal-mode
temperatures without strong effect of the non-Markovian memory kernel. The
mode temperatures strongly depend on the temperature profile of the heat bath
and the boundary conditions, while the non-equilibrium magnon density
distribution can be described only by introducing a chemical potential.
Gradients of magnetic fields cause Wannier-Zeeman magnon localization that
should be observable in the transverse or non-local spin Seebeck effect on
magnetic insulators with strong magnetocrystalline anisotropies such as
(YBi)$_{3}$(FeGa)$_{5}$O$_{12}$. Magnon-magnon interactions can to leading
order be captured by increased mode temperatures. Our generic results shed
light on the fundamental concept of temperature and are applicable to many
disciplines beyond spintronics.

\section*{ACKNOWLEDGMENTS}

This work is supported by the National Natural Science Foundation of China
(NSFC) under Grant No. 11604041, the National Key Research Development Program
under Contract No. 2016YFA0300801, the National Thousand-Young-Talent Program
of China, the DFG Priority Programme 1538 ``Spin-Caloric
Transport\textquotedblright, the NWO, EU FP7 ICT Grant No. 612759 InSpin, and
Grant-in-Aid for Scientific Research (Grant Nos. 25247056, 25220910, and 26103006).

\appendix

\section{Symmetric tridiagonal matrix \v{Q}}

Here we consider the effect of boundary conditions on the canonical $\left(
N+1\right)  \times\left(  N+1\right)  $ matrix $\check{Q}=Q+\left(
\varepsilon/J\right)  $diag$\left\{  n\right\}  $ for the $n=0,1,2,\cdots,N$
spin chain with nearest-neighbor exchange coupling $J$. $Q$ is diagonalized by
a matrix $P,$ i.e., $P^{-1}\check{Q}P=\mathrm{diag}\left\{  \lambda
_{k}\right\}  ,$ which must be orthogonal: $P^{-1}=P^{\text{T}}.$ We first
consider the case of homogeneous magnetic fields $\left(  \varepsilon=0,\text{
so }\check{Q}=Q\right)  $ for different boundary conditions

\textbf{Case I}: For free boundaries at the ends
\begin{equation}
Q=\left(
\begin{array}
[c]{ccccccc}%
1 & -1 & 0 &  &  & \cdots & 0\\
-1 & 2 & -1 & 0 &  &  & \vdots\\
0 & -1 & 2 & -1 &  &  & \\
&  & \ddots & \ddots & \ddots &  & \\
&  &  & -1 & 2 & -1 & 0\\
\vdots &  &  &  & -1 & 2 & -1\\
0 & \cdots &  &  & 0 & -1 & 1
\end{array}
\right)  \label{Q-I}%
\end{equation}
has eigenvalues
\begin{equation}
\lambda_{k}=2\left(  1-\cos\frac{k\pi}{N+1}\right)  ,\text{ \ \ }%
\end{equation}
with $k=0,1,2,\cdots,N\ $and eigenvectors
\begin{equation}
\mathbf{v}_{k}=\left[  \cos\frac{k\pi}{2\left(  N+1\right)  },\cos\frac{3k\pi
}{2\left(  N+1\right)  },\cdots,\cos\frac{\left(  2N+1\right)  k\pi}{2\left(
N+1\right)  }\right]  ^{\text{T}}\
\end{equation}
that can be normalized as
\begin{equation}
\mathbf{u}_{k}=\left\{
\begin{array}
[c]{c}%
\frac{1}{\sqrt{N+1}}\mathbf{v}_{k},\text{ \ \ }k=0\\
\sqrt{\frac{2}{N+1}}\mathbf{v}_{k},\text{ \ }k\neq0
\end{array}
,\right.
\end{equation}
leading to the orthogonal matrix $P$
\begin{equation}
P_{nk}=\left\{
\begin{array}
[c]{c}%
\frac{1}{\sqrt{N+1}},\text{ \ \ \ \ \ \ \ \ \ \ \ \ \ }k=0\\
\sqrt{\frac{2}{N+1}}\cos\frac{\left(  2n+1\right)  k\pi}{2\left(  N+1\right)
},\text{ \ }k\neq0
\end{array}
.\right.
\end{equation}
The temperature matrix defined as
\begin{equation}
\mathcal{T}_{kk^{\prime}}=\sum_{n=0}^{N}P_{nk}P_{nk^{\prime}}T_{n}%
\end{equation}
has diagonal elements
\begin{equation}
\mathcal{T}_{kk}=\left\{
\begin{array}
[c]{c}%
\bar{T},\text{ \ \ \ \ \ \ \ \ \ \ \ \ \ \ \ \ \ \ \ \ \ \ \ \ \ }k=0\\
\bar{T}\left[  1+\frac{\sum_{n=0}^{N}T_{n}\cos\frac{\left(  2n+1\right)  k\pi
}{N+1}}{\sum_{n=0}^{N}T_{n}}\right]  ,\text{ \ }k\neq0
\end{array}
,\right.
\end{equation}
with $\bar{T}=\sum_{n=0}^{N}T_{n}/\left(  N+1\right)  .$ At equilibrium, we
recover $\mathcal{T}_{kk}=\bar{T}$ $\forall k,$ since $\sum_{n=0}^{N}\cos
\frac{\left(  2n+1\right)  k\pi}{N+1}=0$ and $T_{n}=\bar{T}$ $\forall n.$

In the limit of very small Gilbert damping, e.g., $\alpha\simeq10^{-5}$ in
YIG, the magnon density can be approximated as $\rho_{M,n}\simeq\sum
_{k}\left(  P_{nk}\right)  ^{2}k_{B}\mathcal{T}_{kk}/\omega_{k}$, which
becomes exact for constant temperatures.\textit{ }$f\left(  \omega,T\right)
=k_{B}T/\left(  \hbar\omega\right)  $ is the Rayleigh\textendash Jeans
distribution function and $\left(  P_{nk}\right)  ^{2}$ the probability to
find a $k$-magnon at site $n$. At equilibrium, i.e., $T_{n}\equiv T$ $\forall
n,$ all magnons share the temperature of the heat bath ($\mathcal{T}%
_{kk^{\prime}}=T\delta_{kk^{\prime}}$) and $\rho_{M,n}=\gamma_{n}T$ with
$\gamma_{n}=\sum_{k}\left(  P_{nk}\right)  ^{2}k_{B}/\omega_{k}.$ This agrees
with the low-temperature expansion of the Watson-Blume-Vineyard formula by
introducing $\gamma_{n}\equiv\beta_{n}/T_{c}$ with the Curie temperature
$T_{c}.$ We thereby derive expressions for a site-dependent critical exponent
$\beta_{n}$. $\gamma_{n}$ becomes a constant in the thermodynamic limit
$\left(  N\rightarrow\infty\right)  $ as shown in the upper-middle inset of
Fig. 3(a). In the present 1D model, we have
\begin{equation}
\frac{\gamma_{n}}{k_{B}/J}=\frac{1}{N+1}\frac{1}{x}+\frac{1}{\pi}\sum
_{k=0}^{N}\frac{1+\cos\frac{\left(  2n+1\right)  k\pi}{N+1}}{x+2\left(
1-\cos\frac{k\pi}{N+1}\right)  }\frac{\pi}{N+1},
\end{equation}
where $x=H/J.$ Its thermodynamic limit is
\begin{align}
\lim_{N\rightarrow\infty}\frac{\gamma_{n}}{k_{B}/J}  &  =\frac{1}{\pi}\int%
_{0}^{\pi}\frac{1}{x+2\left(  1-\cos y\right)  }dy\nonumber\\
&  =\frac{1}{\sqrt{x\left(  4+x\right)  }}.
\end{align}
We therefore obtain $\lim_{N\rightarrow\infty}\gamma_{n}=k_{B}/\sqrt{H\left(
H+4J\right)  }.$

\textbf{Case II}: For fixed (pinned) boundaries at the two ends$,$ the number
of spins is effectively reduced to $N-1$ and
\begin{equation}
Q=\left(
\begin{array}
[c]{ccccccc}%
2 & -1 & 0 &  &  & \cdots & 0\\
-1 & 2 & -1 &  &  &  & \vdots\\
0 & -1 & 2 & -1 &  &  & \\
&  & \ddots & \ddots & \ddots &  & \\
&  &  & -1 & 2 & -1 & \\
\vdots &  &  &  & -1 & 2 & -1\\
0 & \cdots &  &  &  & -1 & 2
\end{array}
\right)
\end{equation}
has eigenvalues
\begin{equation}
\lambda_{k}=2\left(  1-\cos\frac{k\pi}{N}\right)  ,\text{ \ }%
\end{equation}
\ with $k=1,2,\cdots,N-1,$ and eigenvectors
\begin{equation}
\mathbf{v}_{k}=\left[  \sin\frac{k\pi}{N},\sin\frac{2k\pi}{N},\cdots,\sin
\frac{\left(  N-1\right)  k\pi}{N}\right]  ^{\text{T}},
\end{equation}
normalized as
\begin{equation}
\mathbf{u}_{k}=\sqrt{\frac{2}{N}}\mathbf{v}_{k},
\end{equation}
and the matrix elements of $P$
\begin{equation}
P_{nk}=\sqrt{\frac{2}{N}}\sin\frac{nk\pi}{N},\text{ \ \ }n=1,2,\cdots,N-1.
\end{equation}
Now
\begin{equation}
\mathcal{T}_{kk}=\frac{N-1}{N}\bar{T}\left[  1-\frac{\sum_{n=1}^{N-1}T_{n}%
\cos\frac{2nk\pi}{N}}{\sum_{n=1}^{N-1}T_{n}}\right]  ,\text{ \ \ }%
k=1,2,\cdots,N-1,
\end{equation}
with $\bar{T}=\sum_{n=1}^{N-1}T_{n}/\left(  N-1\right)  .$ Since $\sum
_{n=1}^{N-1}\cos\frac{2nk\pi}{N}=-1,$ we again recover $\mathcal{T}_{kk}%
=\bar{T}$ $\forall k$ at equilibrium.

\textbf{Case III}: For fixed amplitude at site $n=0$ and free amplitude at
site $n=N,$ the number of spins is $N$. The $N\times N$ matrix
\begin{equation}
Q=\left(
\begin{array}
[c]{ccccccc}%
2 & -1 & 0 &  &  & \cdots & 0\\
-1 & 2 & -1 & 0 &  &  & \vdots\\
0 & -1 & 2 & -1 &  &  & \\
&  & \ddots & \ddots & \ddots &  & \\
&  &  & -1 & 2 & -1 & 0\\
\vdots &  &  &  & -1 & 2 & -1\\
0 & \cdots &  &  & 0 & -1 & 1
\end{array}
\right)
\end{equation}
has eigenvalues
\begin{equation}
\lambda_{k}=2\left(  1-\cos\frac{2k-1}{2N+1}\pi\right)  ,\text{ \ \ }%
k=1,2,\cdots,N,
\end{equation}
with $k=1,2,\cdots,N,$ and eigenvectors
\begin{equation}
\mathbf{v}_{k}=\left[  \sin\frac{2k-1}{2N+1}\pi,\sin\frac{2\left(
2k-1\right)  }{2N+1}\pi,\cdots,\sin\frac{N\left(  2k-1\right)  }{2N+1}%
\pi\right]  ^{\text{T}}%
\end{equation}
that can be normalized as $\mathbf{u}_{k}=2\mathbf{v}_{k}/\sqrt{2N+1}$ and
matrix elements
\begin{equation}
P_{nk}=\frac{2}{\sqrt{2N+1}}\sin\frac{n\left(  2k-1\right)  }{2N+1}\pi,\text{
\ \ }n=1,2,\cdots,N.
\end{equation}
Now
\begin{equation}
\mathcal{T}_{kk}=\frac{2N}{2N+1}\bar{T}\left[  1-\frac{\sum_{n=1}^{N}T_{n}%
\cos\frac{2n\left(  2k-1\right)  }{2N+1}\pi}{\sum_{n=1}^{N}T_{n}}\right]
,\text{ \ \ }k=1,2,\cdots,N,
\end{equation}
with $\bar{T}=\sum_{n=1}^{N}T_{n}/N.$ In this case $\sum_{n=1}^{N}\cos
\frac{2n\left(  2k-1\right)  }{2N+1}\pi=-1/2,$ and again we recover
$\mathcal{T}_{kk}=\bar{T}$ $\forall k$ at equilibrium.

\begin{figure}[ptbh]
\begin{centering}
\includegraphics[width=8.5cm]{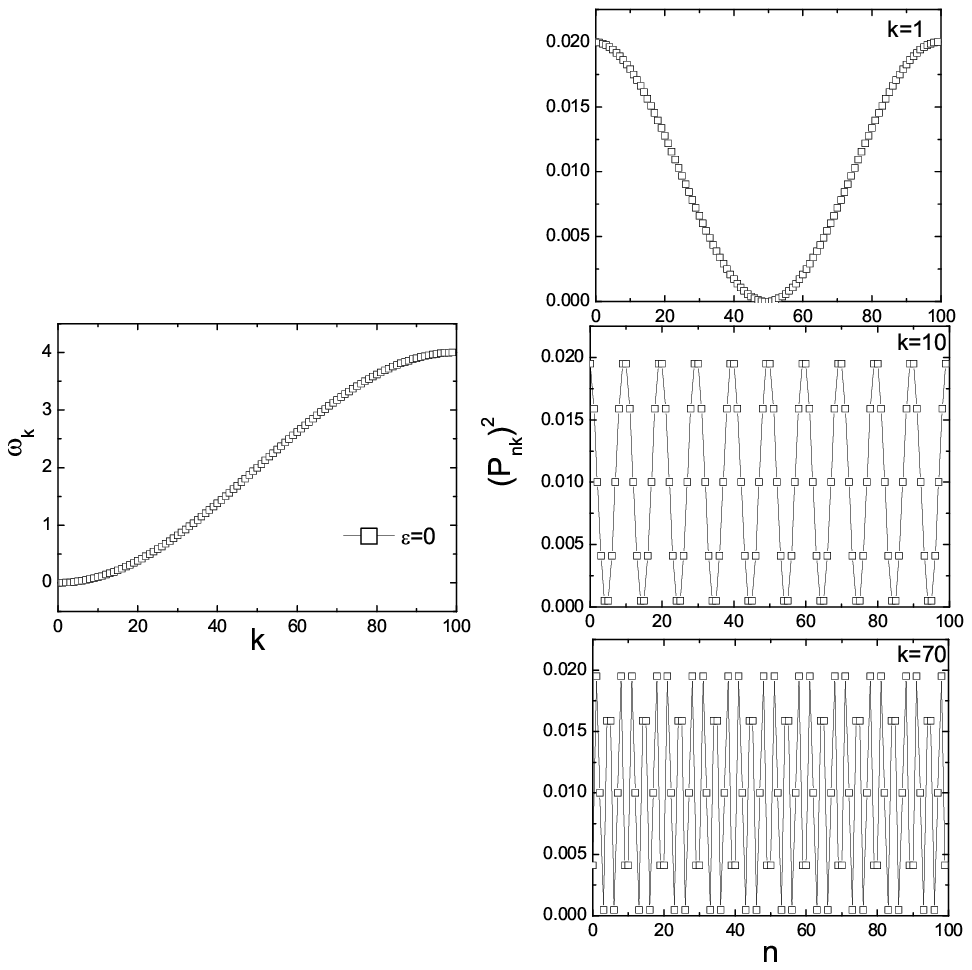}
\par\end{centering}
\caption{(Color online) Magnon dispersion and wave functions without field
gradients.}%
\end{figure}

In the presence of finite field gradients, the matrix $\check{Q}$ generally
cannot be diagonalized analytically. Here, we are interested in the limit of
large magnetic field gradients, i.e., $\left\vert \varepsilon/J\right\vert
\gg1.$ With free boundary conditions, we obtain by perturbation theory
\begin{align}
\lambda_{0}  &  =1,\text{ \ \ \ \ \ \ \ \ \ \ \ \ \ }k=0\nonumber\\
\lambda_{k}  &  =2+\frac{\varepsilon}{J}k,\text{ \ }1\leqslant k\leqslant
N-1\\
\lambda_{N}  &  =1+\frac{\varepsilon}{J}N,\text{ \ \ \ \ \ \ }k=N\nonumber
\end{align}
and
\begin{equation}
P=I_{\left(  N+1\right)  \times\left(  N+1\right)  }\text{ or }P_{nk}%
=\delta_{nk}.
\end{equation}
Correspondingly, the eigenfrequency of the $k$-th mode is
\begin{align}
\omega_{0}  &  =H+J,\text{ \ \ \ \ \ \ \ \ \ \ \ \ \ }k=0\nonumber\\
\omega_{k}  &  =H+2J+\varepsilon k,\text{ \ }1\leqslant k\leqslant N-1\\
\omega_{N}  &  =H+J+\varepsilon N.\text{ \ \ \ \ \ \ }k=N.\nonumber
\end{align}
The spectrum is no longer a trigonometric function of wave number but forms a
Wannier-Zeeman ladder. The temperature matrix
\begin{align}
\mathcal{T}_{kk^{\prime}}  &  =\sum_{n=0}^{N}P_{nk}P_{nk^{\prime}}T_{n}%
=\sum_{n=0}^{N}\delta_{nk}\delta_{nk^{\prime}}T_{n}\nonumber\\
&  =\delta_{kk^{\prime}}T_{k}%
\end{align}
is now diagonal. The mangons are now Wannier-Zeeman localized to the unit cell
rendering the spin chain insulating for spin and energy currents. This can be
illustrated in small damping/Markovian limit with magnonic spin-current
\begin{equation}
j_{M,n}^{z}=J\sum_{k\neq k^{\prime}}P_{nk}P_{\left(  n-1\right)  k^{\prime}%
}k_{B}\mathcal{T}_{kk^{\prime}}\mathcal{F}\left(  \alpha,\omega_{k}%
,\omega_{k^{\prime}}\right)  ,
\end{equation}
where $\mathcal{F}=4\alpha\left(  \omega_{k}-\omega_{k^{\prime}}\right)
/\left[  \alpha^{2}\left(  \omega_{k}+\omega_{k^{\prime}}\right)  ^{2}+\left(
\omega_{k}-\omega_{k^{\prime}}\right)  ^{2}\right]  $ is an anti-symmetric
Lorentzian, that vanishes for a diagonal temperature matrix. The associated
magnon density
\begin{equation}
\rho_{M,n}=\frac{1}{2}\left\langle \psi_{n}^{\ast}\psi_{n}\right\rangle
=\frac{k_{B}T_{n}}{\omega_{n}}%
\end{equation}
indicates local equilibrium. \begin{figure}[ptbh]
\begin{centering}
\includegraphics[width=8.5cm]{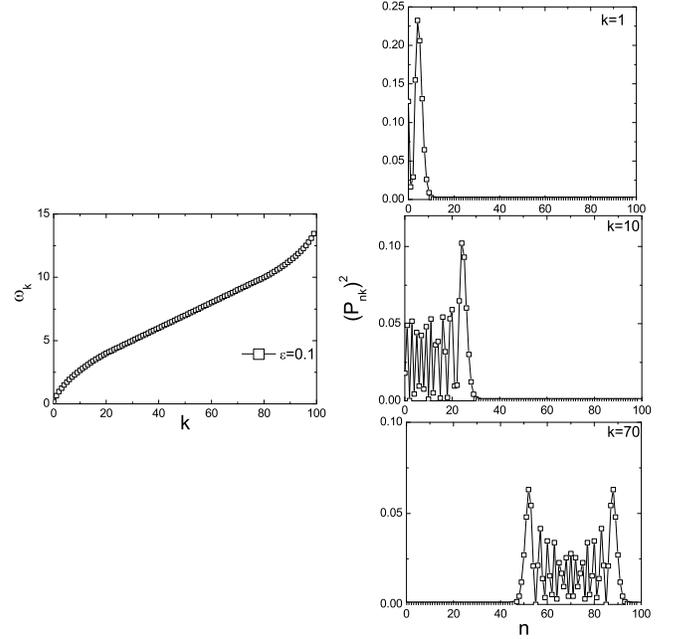}
\par\end{centering}
\caption{(Color online) Magnon dispersion and wave functions with a field
gradient $\varepsilon=0.1$.}%
\end{figure}

In the following, we present numerical calculations for different field
gradients in order to illustrate the transition from propagation Bloch to
localized Wannier-Zeeman states by increasing $\varepsilon.$ Here we adopt
$J=1$, $H=0$ (its value only shifts the magnon band gap), and consider free
boundary conditions. \begin{figure}[ptbh]
\begin{centering}
\includegraphics[width=8.5cm]{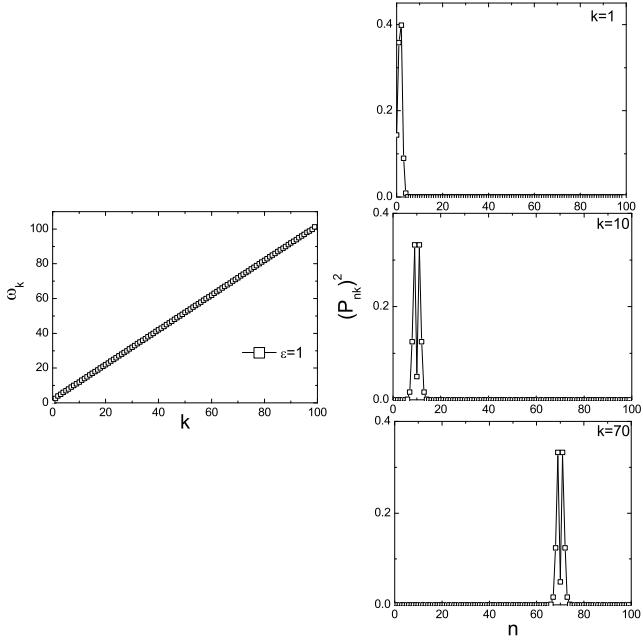}
\par\end{centering}
\caption{(Color online) Magnon dispersion and wave functions with a field
gradient $\varepsilon=1$.}%
\end{figure}

Figure 5 shows the results without field gradients. The magnon dispersion is a
cosine function. The magnon wave functions are spreading Bloch states.

Figure 6 shows the results at $\varepsilon=0.1$. The magnon dispersion is
starting to deviate from the cosine function. The magnon wave functions are localized.

Figure 7 shows the results at $\varepsilon=1$. The magnon dispersion becomes
linear. The magnon wave functions are more localized. \begin{figure}[ptbh]
\begin{centering}
\includegraphics[width=8.5cm]{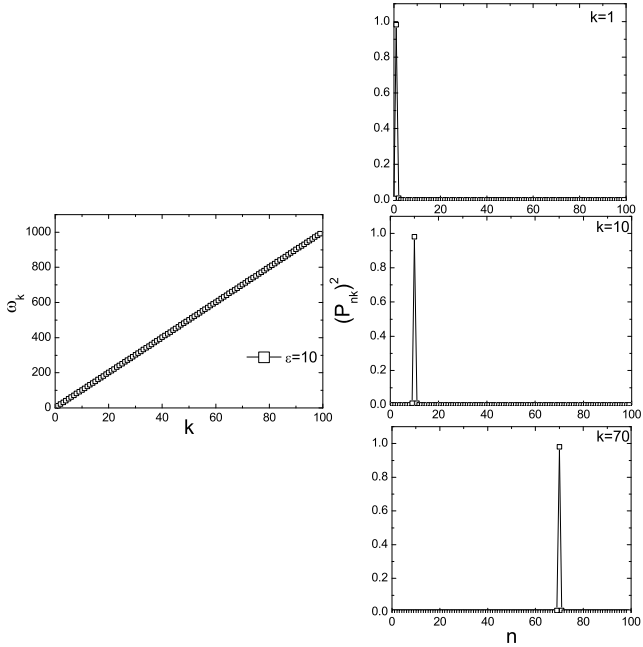}
\par\end{centering}
\caption{(Color online) Magnon dispersion and wave functions with a field
gradient $\varepsilon=10$.}%
\end{figure}

Figure 8 shows the results at $\varepsilon=10$. The magnon dispersion is
linear with strongly localized wave functions. The localization length is
close to a lattice constant. Figures 6-8 show that in the valleys of an
inhomogeneous magnetic field distribution only low-energy magnons contribute,
since high-energy magnons are localized to the hills. The case is opposite in
the high-field side that only high-energy magnons contribute, since low-energy
magnons are localized in the low-field side. \begin{figure}[ptbh]
\begin{centering}
\includegraphics[width=8cm]{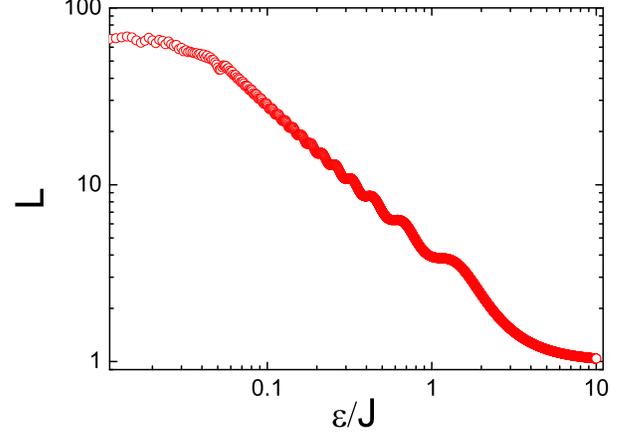}
\par\end{centering}
\caption{(Color online) Localization length $L$ as a function of the field
gradient $\varepsilon/J$.}%
\end{figure}The magnon localization length
\begin{equation}
L\left(  \varepsilon/J\right)  =\frac{1}{\sum_{n=0}^{N}\left(  P_{nk}\right)
^{4}}.
\end{equation}
is plotted in Figure 9 as a function of the field-gradient.

\section{Perturbation Theory}

In this section, we present a perturbative solution of the stochastic
nonlinear equation including the interaction term $\nu$ for arbitrary field
gradients. We expand the normal modes as
\begin{equation}
\phi_{k}\left(  t\right)  =\phi_{k,0}\left(  t\right)  +\nu\phi_{k,1}\left(
t\right)  +\nu^{2}\phi_{k,2}\left(  t\right)  +\cdots,
\end{equation}
and
\begin{equation}
\dot{\phi}_{k}\left(  t\right)  =\dot{\phi}_{k,0}\left(  t\right)  +\nu
\dot{\phi}_{k,1}\left(  t\right)  +\nu^{2}\dot{\phi}_{k,2}\left(  t\right)
+\cdots.
\end{equation}
Keeping only only first-order terms, \begin{widetext}
\[
\left(i+\alpha\right)\left(\dot{\phi}_{k,0}+\nu\dot{\phi}_{k,1}\right)=-\omega_{k}\left(\phi_{k,0}+\nu\phi_{k,1}\right)+\nu\sum_{k_{1},k_{2},k_{3}}I_{k,k_{1},k_{2},k_{3}}\phi_{k_{1},0}^{\ast}\phi_{k_{2},0}\phi_{k_{3},0}+\zeta_{k}\left(t\right).
\]
We therefore obtain
\begin{align}
\text{zero-order} & \text{: }\left(i+\alpha\right)\dot{\phi}_{k,0}=-\omega_{k}\phi_{k,0}+\zeta_{k}\left(t\right),\\
\text{first-order} & \text{: }\left(i+\alpha\right)\dot{\phi}_{k,1}=-\omega_{k}\phi_{k,1}+\sum_{k_{1},k_{2},k_{3}}I_{k,k_{1},k_{2},k_{3}}\phi_{k_{1},0}^{\ast}\phi_{k_{2},0}\phi_{k_{3},0}.
\end{align}
The stationary solution of the zero-order equation is
\begin{equation}
\phi_{k,0}\left(t\right)=\frac{1}{i+\alpha}\int_{-\infty}^{t}dt^{\prime}\exp\left[-\frac{\omega_{k}}{i+\alpha}\left(t-t^{\prime}\right)\right]\zeta_{k}\left(t^{\prime}\right),
\end{equation}
and that for the first-order one is
\begin{equation}
\phi_{k,1}\left(t\right)=\frac{1}{i+\alpha}\int_{-\infty}^{t}dt^{\prime}\exp\left[-\frac{\omega_{k}}{i+\alpha}\left(t-t^{\prime}\right)\right]\sum_{k_{1},k_{2},k_{3}}I_{k,k_{1},k_{2},k_{3}}\phi_{k_{1},0}^{\ast}\left(t^{\prime}\right)\phi_{k_{2},0}\left(t^{\prime}\right)\phi_{k_{3},0}\left(t^{\prime}\right).
\end{equation}
The quantity we aim to evaluate is
\begin{equation}
\frac{\omega_{k}}{2}\left\langle \phi_{k}^{\ast}\left(t\right)\phi_{k}\left(t\right)\right\rangle =\frac{\omega_{k}}{2}\left\langle \phi_{k,0}^{\ast}\left(t\right)\phi_{k,0}\left(t\right)\right\rangle +\nu\omega_{k}\operatorname{Re}\left\langle \phi_{k,0}^{\ast}\left(t\right)\phi_{k,1}\left(t\right)\right\rangle .\label{ModeTemp1}
\end{equation}
The first term in the right-hand side of the above equation is simply
$k_{B}\mathcal{T}_{kk},$ while the second term is
\[
\left\langle \phi_{k,0}^{\ast}\left(t\right)\phi_{k,1}\left(t\right)\right\rangle =\frac{1}{i+\alpha}\int_{-\infty}^{t}dt^{\prime}\exp\left[-\frac{\omega_{k}}{i+\alpha}\left(t-t^{\prime}\right)\right]\sum_{k_{1},k_{2},k_{3}}I_{k,k_{1},k_{2},k_{3}}\left\langle \phi_{k_{1},0}^{\ast}\left(t^{\prime}\right)\phi_{k_{2},0}\left(t^{\prime}\right)\phi_{k_{3},0}\left(t^{\prime}\right)\phi_{k,0}^{\ast}\left(t\right)\right\rangle ,
\]
where correlation
\begin{align*}
& \left\langle \phi_{k_{1},0}^{\ast}\left(t^{\prime}\right)\phi_{k_{2},0}\left(t^{\prime}\right)\phi_{k_{3},0}\left(t^{\prime}\right)\phi_{k,0}^{\ast}\left(t\right)\right\rangle \\
& =\frac{1}{\left(1+\alpha^{2}\right)^{2}}\int_{-\infty}^{t^{\prime}}dt^{\prime\prime}\int_{-\infty}^{t^{\prime}}dt^{\prime\prime\prime}\int_{-\infty}^{t^{\prime}}dt^{\prime\prime\prime\prime}\int_{-\infty}^{t}dt^{\prime\prime\prime\prime\prime}\exp\left[-\frac{\omega_{k_{1}}}{-i+\alpha}\left(t^{\prime}-t^{\prime\prime}\right)-\frac{\omega_{k_{2}}}{i+\alpha}\left(t^{\prime}-t^{\prime\prime\prime}\right)-\frac{\omega_{k_{3}}}{i+\alpha}\left(t^{\prime}-t^{\prime\prime\prime\prime}\right)-\frac{\omega_{k}}{-i+\alpha}\left(t-t^{\prime\prime\prime\prime\prime}\right)\right]\\
& \times\left\langle \zeta_{k_{1}}^{\ast}\left(t^{\prime\prime}\right)\zeta_{k_{2}}\left(t^{\prime\prime\prime}\right)\zeta_{k_{3}}\left(t^{\prime\prime\prime\prime}\right)\zeta_{k}^{\ast}\left(t^{\prime\prime\prime\prime\prime}\right)\right\rangle ,
\end{align*}
By Isserlis' (or Wick's) theorem, we have
\begin{align*}
& \left\langle \zeta_{k_{1}}^{\ast}\left(t^{\prime\prime}\right)\zeta_{k_{2}}\left(t^{\prime\prime\prime}\right)\zeta_{k_{3}}\left(t^{\prime\prime\prime\prime}\right)\zeta_{k}^{\ast}\left(t^{\prime\prime\prime\prime\prime}\right)\right\rangle =\left\langle \zeta_{k_{1}}^{\ast}\left(t^{\prime\prime}\right)\zeta_{k_{2}}\left(t^{\prime\prime\prime}\right)\right\rangle \left\langle \zeta_{k_{3}}\left(t^{\prime\prime\prime\prime}\right)\zeta_{k}^{\ast}\left(t^{\prime\prime\prime\prime\prime}\right)\right\rangle +\left\langle \zeta_{k_{1}}^{\ast}\left(t^{\prime\prime}\right)\zeta_{k_{3}}\left(t^{\prime\prime\prime\prime}\right)\right\rangle \left\langle \zeta_{k_{2}}\left(t^{\prime\prime\prime}\right)\zeta_{k}^{\ast}\left(t^{\prime\prime\prime\prime\prime}\right)\right\rangle \\
& =\left(4\alpha k_{B}\right)^{2}\left[\mathcal{T}_{kk_{3}}\mathcal{T}_{k_{1}k_{2}}\delta\left(t^{\prime\prime}-t^{\prime\prime\prime}\right)\delta\left(t^{\prime\prime\prime\prime}-t^{\prime\prime\prime\prime\prime}\right)+\mathcal{T}_{kk_{2}}\mathcal{T}_{k_{1}k_{3}}\delta\left(t^{\prime\prime}-t^{\prime\prime\prime\prime}\right)\delta\left(t^{\prime\prime\prime}-t^{\prime\prime\prime\prime\prime}\right)\right],
\end{align*}
where we only keep the non-zero terms. After straightforward substitutions
\[
\left\langle \phi_{k,0}^{\ast}\left(t\right)\phi_{k,1}\left(t\right)\right\rangle =\frac{\left(4\alpha k_{B}\right)^{2}\left(-i+\alpha\right)}{\alpha\omega_{k}}\sum_{k_{1},k_{2},k_{3}}I_{k,k_{1},k_{2},k_{3}}\frac{\mathcal{T}_{kk_{3}}\mathcal{T}_{k_{1}k_{2}}}{\left[\omega_{k_{1}}\left(i+\alpha\right)+\omega_{k_{2}}\left(-i+\alpha\right)\right]\left[\omega_{k}\left(i+\alpha\right)+\omega_{k_{3}}\left(-i+\alpha\right)\right]}.
\]
The perturbative mode temperature (\ref{ModeTemp1}) is thus given
by
\begin{align}
& \frac{\omega_{k}}{2}\left\langle \phi_{k}^{\ast}\left(t\right)\phi_{k}\left(t\right)\right\rangle \nonumber \\
& =k_{B}\mathcal{T}_{kk}+16\nu\sum_{k_{1},k_{2},k_{3}}I_{k,k_{1},k_{2},k_{3}}\frac{\alpha^{2}\left(k_{B}\mathcal{T}_{kk_{3}}\right)\left(k_{B}\mathcal{T}_{k_{1}k_{2}}\right)\left[\left(-3+\alpha^{2}\right)\omega_{k}\omega_{k_{1}}+\left(1+\alpha^{2}\right)\left(\omega_{k}\omega_{k_{2}}+\omega_{k_{1}}\omega_{k_{3}}+\omega_{k_{2}}\omega_{k_{3}}\right)\right]}{\left[\left(\omega_{k_{1}}-\omega_{k_{2}}\right)^{2}+\alpha^{2}\left(\omega_{k_{1}}+\omega_{k_{2}}\right)^{2}\right]\left[\left(\omega_{k}-\omega_{k_{3}}\right)^{2}+\alpha^{2}\left(\omega_{k}+\omega_{k_{3}}\right)^{2}\right]}.\label{Firstorder}
\end{align}
\end{widetext}In the limit of a very strong Wannier-Zeeman localization, i.e.,
$P_{nk}=\delta_{nk},P_{nk_{1}}=\delta_{nk_{1}},P_{nk_{2}}=\delta_{nk_{2}},$
and $P_{nk_{3}}=\delta_{nk_{3}}$,
\begin{equation}
I_{k,k_{1},k_{2},k_{3}}=\sum_{n}P_{nk}P_{nk_{1}}P_{nk_{2}}P_{nk_{3}}%
=\delta_{kk_{1}}\delta_{kk_{2}}\delta_{kk_{3}},
\end{equation}
which implies absence of mode coupling. The above mode temperature
(\ref{Firstorder}) is then modified to
\[
\frac{\omega_{k}}{2}\left\langle \phi_{k}^{\ast}\left(  t\right)  \phi
_{k}\left(  t\right)  \right\rangle =k_{B}\mathcal{T}_{kk}\left(  1+\frac{4\nu
k_{B}\mathcal{T}_{kk}}{\omega_{k}^{2}}\right)  .
\]

In the limit of a very weak Gilbert damping, only the \textit{trivial}
resonance terms, i.e., $\omega_{k}=\omega_{k_{3}}$ and $\omega_{k_{1}}%
=\omega_{k_{2}},$ in Eq. (\ref{Firstorder}) survive. We thus have
\[
\frac{\omega_{k}}{2}\left\langle \phi_{k}^{\ast}\left(  t\right)  \phi
_{k}\left(  t\right)  \right\rangle =k_{B}\mathcal{T}_{kk}+4\nu\sum_{k_{1}%
}I_{k,k_{1},k_{1},k}\frac{\left(  k_{B}\mathcal{T}_{kk}\right)  \left(
k_{B}\mathcal{T}_{k_{1}k_{1}}\right)  }{\omega_{k}\omega_{k_{1}}}.
\]
Higher-order perturbation calculations are straightforward if necessary.

\section{Spin Monomer}

We implement numerical calculations for a single spin (spin monomer) in
contact with a thermal bath corresponding to either an isolated classical
atomic moment or a strongly localized normal mode in $k$ space. The equation
of motion including the magnon interaction is simplified to
\begin{equation}
\left(  i+\alpha\right)  \frac{d\phi}{dt}=-\omega\phi+\nu\left\vert
\phi\right\vert ^{2}\phi+\zeta\left(  t\right)  ,
\end{equation}
where we omitted subscripts. Here source term $\zeta\left(  t\right)  =\xi
_{1}\left(  t\right)  +i\xi_{2}\left(  t\right)  $ is the complex noise
defined in the main text, with two real-valued Gaussian white noise sources
(Wiener process) $\xi_{1}\left(  t\right)  $ and $\xi_{2}\left(  t\right)  $.

\begin{figure}[ptbh]
\begin{centering}
\includegraphics[width=8cm]{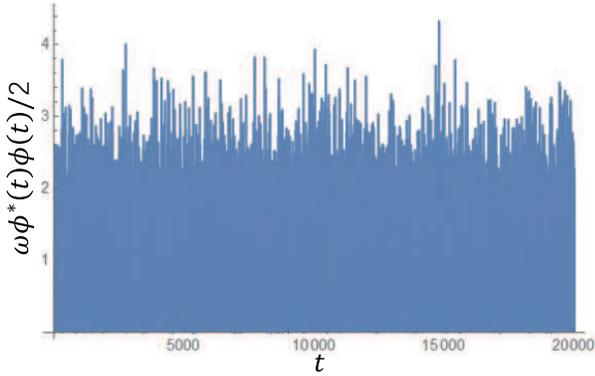}
\par\end{centering}
\caption{(Color online) Time evolution of function $\omega\phi^{\ast}\left(
t\right)  \phi\left(  t\right)  /2$ in a spin monomer driven by a stochastic
white noise.}%
\end{figure}

Figure 10 shows the dynamics of the function $\omega\phi^{\ast}\left(
t\right)  \phi\left(  t\right)  /2.$ We simulate $2\times10^{6}$ steps with a
time step $0.01$ for the time evolution. In numerical calculations, we use
parameters $\omega=k_{B}=\alpha=1,T=1,$ and $\nu=-0.5$. The Ito interpretation
is adopted when integrating the above stochastic differential equation.

\begin{figure}[ptbh]
\begin{centering}
\includegraphics[width=8cm]{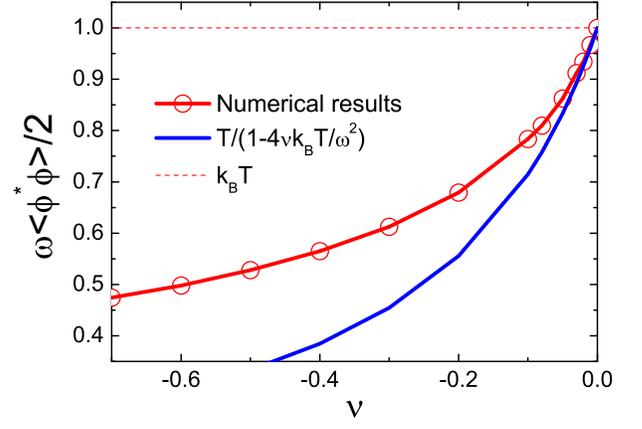}
\par\end{centering}
\caption{(Color online) Renormalization of mode temperature in a spin monomer,
tuned by the strength of nonlinearity $\nu$.}%
\end{figure}

The time-average of $\omega\phi^{\ast}\left(  t\right)  \phi\left(  t\right)
/2$ represents the temperature of the (single) normal mode. Numerical
simulations for every $\nu$ are repeated $20$ times in order to suppress the
statistical error (Figure 10 is just one of them at $\nu=-0.5$). Figure 11
shows the renormalized temperature of the normal mode as a function of the
nonlinearity strength $\nu$. It demonstrates that an increasing nonlinearity
increases the temperature of the mode. In the regime of weak nonlinearity
$\left(  \left\vert \nu\right\vert \leqslant0.02\right)  $ the numerical
results compare very well with the analytical formula.

\section{Spin Dimer}

We implement numerical calculations on a spin dimer model contacting with two
thermal baths with different temperatures. Under free boundary conditions, the
$2\times2$ matrix $\check{Q}$ is
\begin{equation}
\check{Q}=\left(
\begin{array}
[c]{cc}%
1 & -1\\
-1 & 1+\varepsilon/J
\end{array}
\right)  .
\end{equation}
In the following, we set $J=1.$ The corresponding diagonal matrix
\begin{equation}
P=\left(
\begin{array}
[c]{cc}%
\frac{\varepsilon+\sqrt{4+\varepsilon^{2}}}{2\sqrt{1+\frac{1}{4}\left(
\varepsilon+\sqrt{4+\varepsilon^{2}}\right)  ^{2}}} & \frac{\varepsilon
-\sqrt{4+\varepsilon^{2}}}{2\sqrt{1+\frac{1}{4}\left(  \varepsilon
-\sqrt{4+\varepsilon^{2}}\right)  ^{2}}}\\
\frac{1}{\sqrt{1+\frac{1}{4}\left(  \varepsilon+\sqrt{4+\varepsilon^{2}%
}\right)  ^{2}}} & \frac{1}{\sqrt{1+\frac{1}{4}\left(  \varepsilon
-\sqrt{4+\varepsilon^{2}}\right)  ^{2}}}%
\end{array}
\right)  .
\end{equation}
has the eigenvalues
\begin{align}
\omega_{0}  &  =H+\frac{2+\varepsilon-\sqrt{4+\varepsilon^{2}}}{2},\\
\omega_{1}  &  =H+\frac{2+\varepsilon+\sqrt{4+\varepsilon^{2}}}{2}.
\end{align}
For $\varepsilon=1$ the equations of motions for the normal modes in the main
text become
\begin{align}
\left(  i+\alpha\right)  \frac{d\phi_{0}}{dt}  &  =-\omega_{0}\phi_{0}%
+\nu\left(  -0.2\left\vert \phi_{0}\right\vert ^{2}\phi_{0}+0.8\left\vert
\phi_{0}\right\vert ^{2}\phi_{1}+0.2\phi_{0}^{\ast}\phi_{1}^{2}\right.
\nonumber\\
&  +\left.  0.4\phi_{0}^{2}\phi_{1}^{\ast}+0.4\phi_{0}\left\vert \phi
_{1}\right\vert ^{2}+0.6\left\vert \phi_{1}\right\vert ^{2}\phi_{1}\right)
\nonumber\\
&  +\zeta_{0}\left(  t\right)  ,\\
\left(  i+\alpha\right)  \frac{d\phi_{1}}{dt}  &  =-\omega_{1}\phi_{1}%
+\nu\left(  0.6\left\vert \phi_{0}\right\vert ^{2}\phi_{0}-0.4\left\vert
\phi_{0}\right\vert ^{2}\phi_{1}+0.4\phi_{0}^{\ast}\phi_{1}^{2}\right.
\nonumber\\
&  -\left.  0.2\phi_{0}^{2}\phi_{1}^{\ast}+0.8\phi_{0}\left\vert \phi
_{1}\right\vert ^{2}+0.2\left\vert \phi_{1}\right\vert ^{2}\phi_{1}\right)
\nonumber\\
&  +\zeta_{1}\left(  t\right)  ,
\end{align}
with
\begin{align}
\zeta_{0}\left(  t\right)   &  =-0.850651\xi_{0}\left(  t\right)
-0.525731\xi_{1}\left(  t\right)  ,\\
\zeta_{1}\left(  t\right)   &  =-0.525731\xi_{0}\left(  t\right)
+0.850651\xi_{1}\left(  t\right)  ,
\end{align}
in which source terms $\xi_{0}\left(  t\right)  =\xi_{01}\left(  t\right)
+i\xi_{02}\left(  t\right)  $ and $\xi_{1}\left(  t\right)  =\xi_{11}\left(
t\right)  +i\xi_{12}\left(  t\right)  $ with Gaussian white noises (Wiener
process) $\xi_{01}\left(  t\right)  ,\xi_{02}\left(  t\right)  ,\xi
_{11}\left(  t\right)  ,$ and $\xi_{12}\left(  t\right)  $.

\begin{figure}[ptbh]
\begin{centering}
\includegraphics[width=8cm]{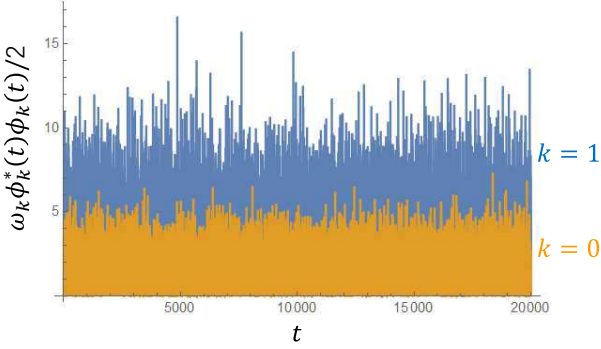}
\par\end{centering}
\caption{(Color online) Time evolution of function $\omega_{k}\phi_{k}^{\ast
}\left(  t\right)  \phi_{k}\left(  t\right)  /2$ for the two normal modes
$(k=0$ and $k=1)$ in a spin dimmer.}%
\end{figure}

Figure 12 shows the dynamics of function $\omega_{k}\phi_{k}^{\ast}\left(
t\right)  \phi_{k}\left(  t\right)  /2$ for $k=0$ and $1.$ We simulate
$2\times10^{6}$ steps with a time step $0.01$ for the time evolution.
Parameters used in the numerical calculations are $H=\varepsilon
=J=k_{B}=\alpha=1,T_{1}=2T_{0}=2,$ and $\nu=-0.6$. Ito interpretation is
adopted to integrate the above stochastic differential equations.

\begin{figure}[ptbh]
\begin{centering}
\includegraphics[width=8cm]{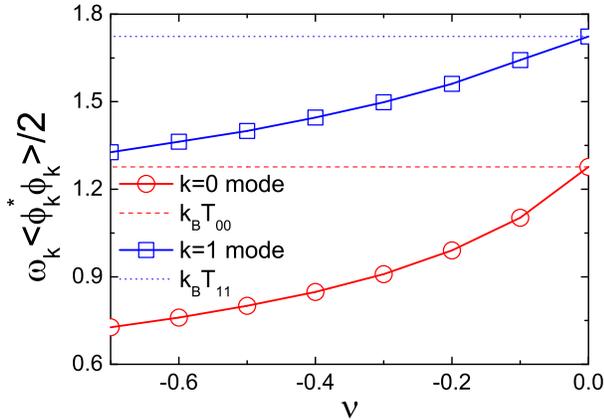}
\par\end{centering}
\caption{(Color online) Renormalization of mode temperatures in a spin dimer,
tuned by the strength of nonlinearity $\nu$. $T_{00}$ and $T_{11}$ represent
the temperatures of normal modes for $k=0$ and $k=1$, respectively, without
nonlinearity.}%
\end{figure}

The time-average of $\omega_{k}\phi_{k}^{\ast}\left(  t\right)  \phi
_{k}\left(  t\right)  /2$ represents the temperature of the normal mode.
Numerical simulations for every $\nu$ are repeated $20$ times (Figure 12 is
just one of them when $\nu=-0.6$). Figure 13 shows the renormalized
temperatures of normal modes as a function of the nonlinearity strength $\nu$.
It demonstrates that an increasing nonlinearity increases the temperature of
all modes.

\section{Spin Trimer}

Numerical calculations of a spin trimer model are presented here. Under free
boundary conditions, the $3\times3$ matrix $\check{Q}$ is
\begin{equation}
\check{Q}=\left(
\begin{array}
[c]{ccc}%
1 & -1 & 0\\
-1 & 2+\varepsilon & -1\\
0 & -1 & 1+2\varepsilon
\end{array}
\right)  ,
\end{equation}
where we assume $J=1.$ Because the analytical form of the eigenvalues and
eigenvector of the above matrix is too complicated, we assign a specific
number to $\varepsilon,$ e.g., $\varepsilon=0.5.$ The corresponding diagonal
matrix then reads
\[
P=\left(
\begin{array}
[c]{ccc}%
-0.313433 & -0.516706 & 0.796727\\
0.796727 & 0.313433 & 0.516706\\
-0.516706 & 0.796727 & 0.313433
\end{array}
\right)  ,
\]
and the eigen values of three normal modes are
\begin{align}
\omega_{0}  &  =H+0.351465,\\
\omega_{1}  &  =H+1.6066,\\
\omega_{2}  &  =H+3.54194.
\end{align}
In the following numerical calculations, we use parameters $H=k_{B}%
=\alpha=1,T_{0}=1,T_{1}=2,T_{2}=3.$ The three eigenfrequencies are then
$\omega_{0}=1.351465,\omega_{1}=2.6066,$ and $\omega_{2}=4.54194.$ The
equations of motions for normal modes become
\begin{align}
\left(  i+\alpha\right)  \frac{d\phi_{0}}{dt}  &  =-\omega_{0}\phi_{0}%
+\nu\left(  0.193548\left\vert \phi_{0}\right\vert ^{2}\phi_{0}%
+0.516129\left\vert \phi_{0}\right\vert ^{2}\phi_{2}\right. \nonumber\\
&  -0.0645161\phi_{0}^{\ast}\phi_{2}^{2}+0.258065\left\vert \phi
_{0}\right\vert ^{2}\phi_{1}\nonumber\\
&  +0.258065\phi_{0}^{\ast}\phi_{1}\phi_{2}-0.129032\phi_{0}^{\ast}\phi
_{1}^{2}\nonumber\\
&  +0.258065\phi_{0}^{2}\phi_{2}^{\ast}-0.129032\phi_{0}\left\vert \phi
_{2}\right\vert ^{2}\nonumber\\
&  +0.483871\left\vert \phi_{2}\right\vert ^{2}\phi_{2}+0.258065\phi_{0}%
\phi_{1}\phi_{2}^{\ast}\nonumber\\
&  -0.387097\phi_{1}\left\vert \phi_{2}\right\vert ^{2}+0.258065\phi_{1}%
^{2}\phi_{2}^{\ast}\nonumber\\
&  +0.129032\phi_{0}^{2}\phi_{1}^{\ast}+0.258065\phi_{0}\phi_{1}^{\ast}%
\phi_{2}\nonumber\\
&  -0.193548\phi_{1}^{\ast}\phi_{2}^{2}-0.258065\phi_{0}\left\vert \phi
_{1}\right\vert ^{2}\nonumber\\
&  +\left.  0.516129\left\vert \phi_{1}\right\vert ^{2}\phi_{2}%
+0.0645161\left\vert \phi_{1}\right\vert ^{2}\phi_{1}\right) \nonumber\\
&  +\zeta_{0}\left(  t\right)  ,\\
\left(  i+\alpha\right)  \frac{d\phi_{1}}{dt}  &  =-\omega_{1}\phi_{1}%
+\nu\left(  0.0645161\left\vert \phi_{0}\right\vert ^{2}\phi_{0}%
+0.258065\left\vert \phi_{0}\right\vert ^{2}\phi_{2}\right. \nonumber\\
&  +0.129032\phi_{0}^{\ast}\phi_{2}^{2}+0.516129\left\vert \phi_{0}\right\vert
^{2}\phi_{1}\nonumber\\
&  -0.258065\phi_{0}^{\ast}\phi_{1}\phi_{2}-0.193548\phi_{0}^{\ast}\phi
_{1}^{2}\nonumber\\
&  +0.129032\phi_{0}^{2}\phi_{2}^{\ast}+0.258065\phi_{0}\left\vert \phi
_{2}\right\vert ^{2}\nonumber\\
&  -0.193548\left\vert \phi_{2}\right\vert ^{2}\phi_{2}-0.258065\phi_{0}%
\phi_{1}\phi_{2}^{\ast}\nonumber\\
&  +0.516129\phi_{1}\left\vert \phi_{2}\right\vert ^{2}+0.0645161\phi_{1}%
^{2}\phi_{2}^{\ast}\nonumber\\
&  +0.258065\phi_{0}^{2}\phi_{1}^{\ast}-0.258065\phi_{0}\phi_{1}^{\ast}%
\phi_{2}\nonumber\\
&  +0.258065\phi_{1}^{\ast}\phi_{2}^{2}-0.387097\phi_{0}\left\vert \phi
_{1}\right\vert ^{2}\nonumber\\
&  +\left.  0.129032\left\vert \phi_{1}\right\vert ^{2}\phi_{2}%
+0.483871\left\vert \phi_{1}\right\vert ^{2}\phi_{1}\right) \nonumber\\
&  +\zeta_{1}\left(  t\right)  ,\\
\left(  i+\alpha\right)  \frac{d\phi_{2}}{dt}  &  =-\omega_{2}\phi_{2}%
+\nu\left(  0.483871\left\vert \phi_{0}\right\vert ^{2}\phi_{0}%
+0.387097\left\vert \phi_{0}\right\vert ^{2}\phi_{2}\right. \nonumber\\
&  +0.258065\phi_{0}^{\ast}\phi_{2}^{2}+0.129032\left\vert \phi_{0}\right\vert
^{2}\phi_{1}\nonumber\\
&  +0.258065\phi_{0}^{\ast}\phi_{1}\phi_{2}+0.258065\phi_{0}^{\ast}\phi
_{1}^{2}\nonumber\\
&  +0.193548\phi_{0}^{2}\phi_{2}^{\ast}+0.516129\phi_{0}\left\vert \phi
_{2}\right\vert ^{2}\nonumber\\
&  -0.0645161\left\vert \phi_{2}\right\vert ^{2}\phi_{2}+0.258065\phi_{0}%
\phi_{1}\phi_{2}^{\ast}\nonumber\\
&  +0.258065\phi_{1}\left\vert \phi_{2}\right\vert ^{2}-0.129032\phi_{1}%
^{2}\phi_{2}^{\ast}\nonumber\\
&  +0.0645161\phi_{0}^{2}\phi_{1}^{\ast}+0.258065\phi_{0}\phi_{1}^{\ast}%
\phi_{2}\nonumber\\
&  +0.129032\phi_{1}^{\ast}\phi_{2}^{2}+0.516129\phi_{0}\left\vert \phi
_{1}\right\vert ^{2}\nonumber\\
&  -\left.  0.258065\left\vert \phi_{1}\right\vert ^{2}\phi_{2}%
-0.193548\left\vert \phi_{1}\right\vert ^{2}\phi_{1}\right) \nonumber\\
&  +\zeta_{2}\left(  t\right)  ,
\end{align}
with
\begin{align}
\zeta_{0}\left(  t\right)   &  =0.796727\xi_{0}\left(  t\right)
+0.516706\xi_{1}\left(  t\right)  +0.313433\xi_{2}\left(  t\right)  ,\\
\zeta_{1}\left(  t\right)   &  =-0.516706\xi_{0}\left(  t\right)
+0.313433\xi_{1}\left(  t\right)  +0.796727\xi_{2}\left(  t\right)  ,\\
\zeta_{2}\left(  t\right)   &  =-0.313433\xi_{0}\left(  t\right)
+0.796727\xi_{1}\left(  t\right)  -0.516706\xi_{2}\left(  t\right)  ,
\end{align}
in which source terms $\xi_{0}\left(  t\right)  =\xi_{01}\left(  t\right)
+i\xi_{02}\left(  t\right)  ,\xi_{1}\left(  t\right)  =\xi_{11}\left(
t\right)  +i\xi_{12}\left(  t\right)  $ and $\xi_{2}\left(  t\right)
=\xi_{21}\left(  t\right)  +i\xi_{22}\left(  t\right)  $ with Gaussian white
noises (Wiener process) $\xi_{01}\left(  t\right)  ,\xi_{02}\left(  t\right)
,\xi_{11}\left(  t\right)  ,\xi_{12}\left(  t\right)  ,\xi_{21}\left(
t\right)  ,$ and $\xi_{22}\left(  t\right)  $. \begin{figure}[ptbh]
\begin{centering}
\includegraphics[width=8cm]{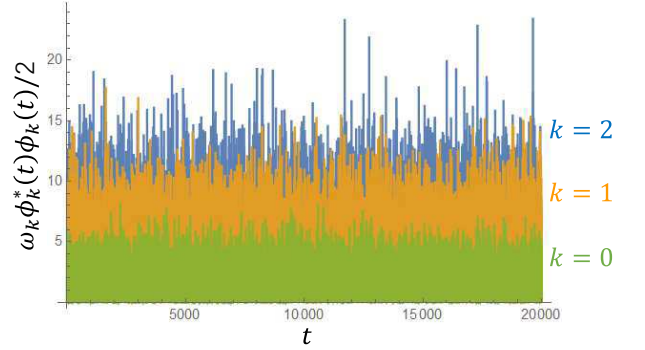}
\par\end{centering}
\caption{(Color online) Time evolution of function $\omega_{k}\phi_{k}^{\ast
}\left(  t\right)  \phi_{k}\left(  t\right)  /2$ for the three normal modes
$(k=0,k=1$ and $k=2)$ in a spin trimmer. The nonlinearity strength is
$\nu=-0.5.$}%
\end{figure}

Figure 14 shows the dynamics of function $\omega_{k}\phi_{k}^{\ast}\left(
t\right)  \phi_{k}\left(  t\right)  /2$ with $k=0,1$ and $2.$ We simulate
$2\times10^{6}$ steps with a time step $0.01$ for the time evolution. Ito
interpretation is adopted to integrate the above stochastic differential equations.

\begin{figure}[ptbh]
\begin{centering}
\includegraphics[width=8cm]{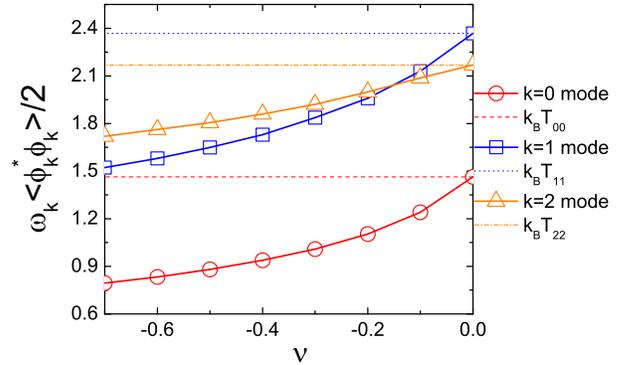}
\par\end{centering}
\caption{(Color online) Renormalization of mode temperatures in a spin trimer,
tuned by the nonlinearity parameter $\nu$. $T_{00},T_{11}$ and $T_{22}$
represent the temperatures of normal modes for $k=0,k=1,$ and $k=2$,
respectively, without nonlinearity.}%
\end{figure}

The time-average of $\omega_{k}\phi_{k}^{\ast}\left(  t\right)  \phi
_{k}\left(  t\right)  /2$ represents the temperature of the normal mode.
Numerical simulations for every $\nu$ are repeated $20$ times in order to
diminish the sample deviation (Figure 14 is one example of them at $\nu
=-0.5$). Figure 15 shows the renormalized temperatures of normal modes as a
function of the nonlinearity strength $\nu$. It demonstrates similar red-shift
behavior as that in spin dimers.


\begin{thebibliography}{99}                                                                                               %


\bibitem {Huang}M. Born and K. Huang, \emph{Dynamical Theory of Crystal
Lattices} (Clarendon Press, Oxford, 1998).

\bibitem {Tolman}R.C. Tolman, A General Theory of Energy Partition with
Applications to Quantum Theory, Phys. Rev. \textbf{11}, 261 (1918).

\bibitem {Callen}T. Callen, \emph{Thermodynamics} 2nd ed. (Wiley, New York, 1985).

\bibitem {Rugh}H.H. Rugh, Dynamical Approach to Temperature, Phys. Rev. Lett.
\textbf{78}, 772 (1997).

\bibitem {Speck}T. Speck and U. Seifert, Restoring a fluctuation-dissipation
theorem in a nonequilibrium steady state, Europhys. Lett. \textbf{74}, 391 (2006).

\bibitem {Huse}R. Nandkishore and D.A. Huse, Many-body Localization and
Thermalization in Quantum Statistical Mechanics, Annu. Rev. Condens. Matter
Phys. \textbf{6}, 15 (2015).

\bibitem {Gerrit}G.E.W. Bauer, E. Saitoh, and B.J. van Wees, Spin
caloritronics, Nat. Mater. \textbf{11}, 391 (2012).

\bibitem {Narayanan}K.R. Narayanan and A.R. Srinivasa, Shannon-entropy-based
nonequilibrium ``entropic\textquotedblright\ temperature of a general
distribution, Phys. Rev. E \textbf{85}, 031151 (2012).

\bibitem {Sasa}S.-ichi Sasa and Y. Yokokura, Thermodynamic Entropy as a
Noether Invariant, Phys. Rev. Lett. \textbf{116}, 140601 (2016).

\bibitem {Alicki}R. Alicki, D. Gelbwaser-Klimovsky, and A. Jenkins, A
thermodynamic cycle for the solar cell, arXiv:1606.03819; R. Alicki and D.
Gelbwaser-Klimovsky, Non-equilibrium quantum heat machines, New J. Phys.
\textbf{17}, 115012 (2015); P. Calabrese, F.H.L. Essler, and M. Fagotti,
Quantum Quench in the Transverse-Field Ising Chain, Phys. Rev. Lett.
\textbf{106}, 227203 (2011).

\bibitem {Miyazaki}K. Miyazaki and K. Seki, Brownian motion of spins
revisited, J. Chem. Phys. \textbf{108}, 7052 (1998).

\bibitem {Brown}W.F. Brown, Jr., Thermal Fluctuations of a Single-Domain
Particle, Phys. Rev. \textbf{130}, 1677 (1963).

\bibitem {Kubo}R. Kubo and N. Hashitsume, Brownian Motion of Spins, Prog.
Theor. Phys. Suppl. \textbf{46}, 210 (1970).

\bibitem {Palacios}J.L. García-Palacios and F.J. Lázaro, Langevin-dynamics
study of the dynamical properties of small magnetic particles, Phys. Rev. B
\textbf{58}, 14937 (1998).

\bibitem {Gilbert}T.L. Gilbert, A phenomenological theory of damping in
ferromagnetic materials, IEEE Trans. Magn. \textbf{40}, 3443 (2004).

\bibitem {Kawabata}A. Kawabata, Brownian Motion of a Classical Spin, Prog.
Theor. Phys. \textbf{48}, 2237 (1972).

\bibitem {Roland}A.A. Khajetoorians, B. Baxevanis, C. Hübner, T. Schlenk, S.
Krause, T.O. Wehling, S. Lounis, A. Lichtenstein, D. Pfannkuche, J. Wiebe, and
R. Wiesendanger, Current-Driven Spin Dynamics of Artificially Constructed
Quantum Magnets, Science \textbf{339}, 55 (2013).

\bibitem {Welton}H.B. Callen and T.A. Welton, Irreversibility and Generalized
Noise, Phys. Rev. \textbf{83}, 34 (1951).

\bibitem {Atxitia}U. Atxitia, O. Chubykalo-Fesenko, R.W. Chantrell, U. Nowak,
and A. Rebei, Ultrafast Spin Dynamics: The Effect of Colored Noise, Phys. Rev.
Lett. \textbf{102}, 057203 (2009).

\bibitem {Peter}A. Rückriegel and P. Kopietz, Rayleigh-Jeans Condensation of
Pumped Magnons in Thin-Film Ferromagnets, Phys. Rev. Lett. \textbf{115},
157203 (2015).

\bibitem {Antropov}V.P. Antropov, V.N. Antonov, L.V. Bekenov, A. Kutepov, and
G. Kotliar, Magnetic anisotropic effects and electronic correlations in MnBi
ferromagnet, Phys. Rev. B \textbf{90}, 054404 (2014).

\bibitem {Sukhov}A. Sukhov, L. Chotorlishvili, A. Ernst, X. Zubizarreta, S.
Ostanin, I. Mertig, E.K.U. Gross, and J. Berakdar, Swift thermal steering of
domain walls in ferromagnetic MnBi stripes, Sci. Rep. \textbf{6}, 24411 (2016).

\bibitem {Zarate}J.M. Ortiz de Zárate and J.V. Sengers, On the Physical Origin
of Long-Ranged Fluctuations in Fuids in Thermal Nonequilibrium States, J.
Stat. Phys. \textbf{115}, 1341 (2004).

\bibitem {Tremblay}A.-M.S. Tremblay, M. Arai, and E.D. Siggia, Fluctuations
about simple nonequilibrium steady states, Phys. Rev. A \textbf{23}, 1451 (1981).

\bibitem {Lorenzo}L. Bertini, A. De Sole, D. Gabrielli, G. Jona-Lasinio, and
C. Landim, Macroscopic fluctuation theory, Rev. Mod. Phys. \textbf{87}, 593 (2015).

\bibitem {Raghavan}S. Raghavan, A. Smerzi, S. Fantoni, and S.R. Shenoy,
Coherent oscillations between two weakly coupled Bose-Einstein condensates:
Josephson effects, $\pi$ oscillations, and macroscopic quantum selft-trapping,
Phys. Rev. A \textbf{59}, 620 (1999).

\bibitem {Joe}S. Geprägs, A. Kehlberger, F.D. Coletta, Z. Qiu, E.-J. Guo, T.
Schulz, C. Mix, S. Meyer, A. Kamra, M. Althammer, H. Huebl, G. Jakob, Y.
Ohnuma, H. Adachi, J. Barker, S. Maekawa, G.E.W. Bauer, E. Saitoh, R. Gross,
S.T.B. Goennenwein, and M. Kläui, Origin of the spin Seebeck effect in
compensated ferrimagnets, Nature Commun. \textbf{7}, 10452 (2016); J. Barker
and G.E.W. Bauer, Thermal Spin Dynamics of Yttrium Iron Garnet, Phys. Rev.
Lett. \textbf{117}, 217201 (2016).

\bibitem {Zotos}A.V. Savin, G.P. Tsironis, and X. Zotos, Thermal conductivity
of a classical one-dimensional spin-phonon system, Phys. Rev. B \textbf{75},
214305 (2007).

\bibitem {Jencic}B. Jen\v{c}i\v{c} and P. Prelovšek, Spin and thermal
conductivity in a classical disordered spin chain, Phys. Rev. B \textbf{92},
134305 (2015).

\bibitem {ShufengZhangprl}S.S.-L. Zhang and S. Zhang, Magnon Mediated Electric
Current Drag Across a Ferromagnetic Insulator Layer, Phys. Rev. Lett.
\textbf{109}, 096603 (2012).

\bibitem {Sanders}D.J. Sanders and D. Walton, Effect of magnon-phonon thermal
relaxation on heat transport by magnons, Phys. Rev. B \textbf{15}, 1489 (1977).

\bibitem {Xiao}J. Xiao, G.E.W. Bauer, K. Uchida, E. Saitoh, and S. Maekawa,
Theory of magnon-driven spin Seebeck effect, Phys. Rev. B \textbf{81}, 214418 (2010).

\bibitem {Konstantin}K.S. Tikhonov, J. Sinova, and A.M. Finkel'stein, Spectral
non-uniform temperature and non-local heat transfer in the spin Seebeck
effect, Nat. Commun. \textbf{4}, 1945 (2013).

\bibitem {Ritzmann}U. Ritzmann, D. Hinzke, and U. Nowak, Propagation of
thermally induced magnonic spin currents, Phys. Rev. B \textbf{89}, 024409 (2014).

\bibitem {Uchida1}K. Uchida, J. Xiao, H. Adachi, J. Ohe, S. Takahashi, J.
Ieda, T. Ota, Y. Kajiwara, H. Umezawa, H. Kawai, G.E.W. Bauer, S. Maekawa, and
E. Saitoh, Spin Seebeck insulator, Nat. Mater. \textbf{9}, 894 (2010).

\bibitem {Uchida2}K. Uchida, H. Adachi, T. Ota, H. Nakayama, S. Maekawa, and
E. Saitoh, Observation of longitudinal spin-Seebeck effect in magnetic
insulators, Appl. Phys. Lett. \textbf{97}, 172505 (2010).

\bibitem {SHuang}S.Y. Huang, X. Fan, D. Qu, Y.P. Chen, W.G. Wang, J. Wu, T.Y.
Chen, J.Q. Xiao, and C.L. Chien, Transport Magnetic Proximity Effects in
Platinum, Phys. Rev. Lett. \textbf{109}, 107204 (2012).

\bibitem {Qu}D. Qu, S.Y. Huang, J. Hu, R. Wu, and C.L. Chien, Intrinsic Spin
Seebeck Effect in Au/YIG, Phys. Rev. Lett. \textbf{110}, 067206 (2013).

\bibitem {Weiler}M. Weiler, M. Althammer, F.D. Czeschka, H. Huebl, M.S.
Wagner, M. Opel, I.-M. Imort, G. Reiss, A. Thomas, R. Gross, and S.T.B.
Goennenwein, Local Charge and Spin Currents in Magnetothermal Landscapes,
Phys. Rev. Lett. \textbf{108}, 106602 (2012).

\bibitem {Ken}K. Uchida, T Kikkawa, A Miura, J Shiomi, and E. Saitoh,
Quantitative Temperature Dependence of Longitudinal Spin Seebeck Effect at
High Temperatures, Phys. Rev. X \textbf{4}, 041023 (2014).

\bibitem {Andreas}A. Kehlberger, U. Ritzmann, D. Hinzke, E.-J. Guo, J. Cramer,
G. Jakob, M.C. Onbasli, D.H. Kim, C.A. Ross, M.B. Jungfleisch, B. Hillebrands,
U. Nowak, and M. Kläui, Length Scale of the Spin Seebeck Effect, Phys. Rev.
Lett. \textbf{115}, 096602 (2015).

\bibitem {Jaworski1}C.M. Jaworski, J. Yang, S. Mack, D.D. Awschalom, J.P.
Heremans, and R.C. Myers, Observation of the spin-Seebeck effect in a
ferromagnetic semiconductor, Nat. Mater. \textbf{9}, 898 (2010).

\bibitem {Jaworski2}C.M. Jaworski, R.C. Myers, E. Johnston-Halperin, and J.P.
Heremans, Giant spin Seebeck effect in a non-magnetic material, Nature
(London) \textbf{487}, 210 (2012).

\bibitem {Zink}A.D. Avery, M.R. Pufall, and B.L. Zink, Observation of the
Planar Nernst Effect in Permalloy and Nickel Thin Films with In-Plane Thermal
Gradients, Phys. Rev. Lett. \textbf{109}, 196602 (2012).

\bibitem {Schmid}M. Schmid, S. Srichandan, D. Meier, T. Kusche, J.-M.
Schmalhorst, M. Vogel, G. Reiss, C. Strunk, and C.H. Back, Transverse Spin
Seebeck Effect versus Anomalous and Planar Nernst Effects in Permalloy Thin
Films, Phys. Rev. Lett. \textbf{111}, 187201 (2013).

\bibitem {Meier}D. Meier, D. Reinhardt, M. van Straaten, C. Klewe, M.
Althammer, M. Schreier, S.T.B. Goennenwein, A. Gupta, M. Schmid, C.H. Back,
J.-M. Schmalhorst, T. Kuschel, and G. Reiss, Longitudinal spin Seebeck effect
contribution in transverse spin Seebeck effect experiments in Pt/YIG and
Pt/NFO, Nat. Commun. \textbf{6}, 8211 (2015).

\bibitem {Shan}J. Shan, L.J. Cornelissen, N. Vlietstra, J.B. Youssef, T.
Kuschel, R.A. Duine, and B.J. van Wees, Influence of yttrium iron garnet
thickness and heater opacity on the nonlocal transport of electrically and
thermally excited magnons, Phys. Rev. B \textbf{94}, 174437 (2016).

\bibitem {Hujun}H. Jiao and G.E.W. Bauer, Spin Backflow and ac Voltage
Generation by Spin Pumping and the Inverse Spin Hall Effect, Phys. Rev. Lett.
\textbf{110}, 217602 (2013).

\bibitem {Kai}K. Chen and S. Zhang, Spin Pumping in the Presence of Spin-Orbit
Coupling, Phys. Rev. Lett. \textbf{114}, 126602 (2015).

\bibitem {Emin}D. Emin and C.F. Hart, Existence of Wannier-Stark localization,
Phys. Rev. B \textbf{36}, 7353 (1987).

\bibitem {Deleeuw}F.H. de Leeuw, Wall velocity in garnet films at high drive
fields, IEEE Trans. Magn. \textbf{13}, 1172 (1977).

\bibitem {Tsang}C. Tsang, C. Bonhote, Q. Dai, H. Do, B. Knigge, Y. Ikeda, Q.
Le, B. Lengsfield, J. Lille, J. Li, S. MacDonald, A. Moser, V. Nayak, R.
Payne, N. Robertson, M. Schabes, N. Smith, K. Takano, P. van der Heijden, W.
Weresin, M. Williams, and M. Xiao, Head challenges for perpendicular recording
at high areal density, IEEE Trans. Magn. \textbf{42}, 145 (2006).

\bibitem {Coey}J.M.D. Coey, New permanent magnets; mangnese compounds, J.
Phys.: Condens. Matter \textbf{26}, 064211 (2014).

\bibitem {Novoselov}K.S. Novoselov, A.K. Geim, S.V. Dubonos, E.W. Hill, and
I.V. Grigorieva, Subatomic movements of a domain wall in the Peierls
potential, Nature (London) \textbf{426}, 812 (2003).

\bibitem {Cherepanov}V. Cherepanov, I. Kolokolov, and V. L'vov, The saga of
YIG: spectra, thermodynamics, interaction and relaxation of magnons in a
complex magnet, Phys. Rep. \textbf{229}, 81 (1993).

\bibitem {Kreisel}A. Kreisel, F. Saulo, L. Bartosch, and P. Kopietz,
Microscopic spin-wave theory for yttrium-iron garnet films, Eur. Phys. J. B
\textbf{71}, 59 (2009).

\bibitem {Rossi}F. Rossi, Bloch oscillations and Wannier-Stark localization in
semiconductor superlattices, in: Theory of Transport Properties of
Semiconductor Nanostructures, edited by E. Schöll, (Spinger, Boston, 1998),
pp. 283\textendash320.

\bibitem {Kosevich}A.M. Kosevich, B.A. Ivanov, and A.S. Kovalev, Magnetic
Solitons, Phys. Rep. \textbf{194}, 117 (1990).

\bibitem {Slavin}A. Slavin and V. Tiberkevich, Nonlinear Auto-Oscillator
Theory of Microwave Generation by Spin-Polarized Current, IEEE Trans. Magn.
\textbf{45}, 1875 (2009).

\bibitem {Simon}S. Borlenghi, S. Iubini, S. Lepri, J. Chico, L. Bergqvist, A.
Delin, and J. Fransson, Energy and magnetization transport in nonequilibrium
macrospin systems, Phys. Rev. E \textbf{92}, 012116 (2015).
\end{thebibliography}
\end{document}